\documentclass[preprint,5p,times,twocolumn,authoryear]{elsarticle}

\usepackage{hyperref}
\usepackage{lineno}

\usepackage{amsmath,amssymb,color,array,natbib,url}

\begin{document}

\begin{frontmatter}
\title{Forbidden mass ranges for shower meteoroids}
\author{Althea V.~Moorhead} \ead{althea.moorhead@nasa.gov}
\address{NASA Meteoroid Environment Office, 
Marshall Space Flight Center, Huntsville, Alabama 35812}

\begin{abstract}
\cite{1979Icar...40....1B} use the parameter $\beta$ to describe the ratio of radiation pressure to gravity experienced by a small particle in the solar system.  The central potential that these particles experience is effectively reduced by a factor of $(1-\beta)$, lowering the escape velocity.  \cite{1979Icar...40....1B} also provide a simple expression for the value of $\beta$ at which particles ejected from a comet follow parabolic orbits and thus leave the solar system; this paper expands on this to calculate critical $\beta$ values that take ejection velocity into account.  We first derive an expression for the speed at which particles are ejected from a comet that is valid at all particle sizes.  We then calculate $\beta$ values for both idealized and real materials to determine which particles will be lost from meteor streams.  We perform these calculations for cometary ejecta leading, trailing, and following the parent comet's nucleus for 10 major meteor showers.   These values bound mass regimes within which meteoroids are ejected from the solar system and therefore cannot contribute to meteor showers.
\end{abstract}

\end{frontmatter}


\section{Introduction}

The orbits of small particles in the solar system are governed by radiative forces as well as gravity.  These forces include radiation pressure, Poynting-Robertson drag, and solar wind pressure. Radiation pressure is the most significant of these and follows an inverse-square law (provided that the Sun is approximated as a point source of radiation).  The ratio of radiation pressure to gravity is thus independent of distance and can be expressed as a dimensionless variable, $\beta$.  
When $\beta = 1$, radiation pressure completely counteracts gravity and particles become unbound from the Sun's gravitational field.  For $\beta < 1$, the potential well remains, but its strength is reduced and the escape speed is lowered. 

A comet with a large orbital eccentricity has an orbital speed at perihelion that is quite close to the nominal local escape speed. The small particles it releases have an even lower escape velocity but are, on average, moving with their parent comet. Thus, these ejected particles may be moving at speeds that exceed their escape velocities \citep{1963JGR....68.2171H,1970JGR....75.3468D}.  This situation likely occurs for many meteoroids: most meteor showers have been linked to either comets or comet-like asteroids (such as 2003 EH1 and 3200 Phaethon), and the sporadic meteor background is also largely attributed to comets \citep{2009Icar..201..295W,2011ApJ...743..129N,2014ApJ...789...25P}.  

The radiation pressure factor, $\beta$, tends to be inversely proportional to diameter for large particles.  However, particles smaller than the radiation wavelength do not absorb photons as readily, and the inverse relationship between $\beta$ and size breaks down at small sizes.  Depending on the optical properties of the ejecta, $\beta$ may be less than one at all sizes \cite[such as basalt; see Fig.~7a of][]{1979Icar...40....1B}.  

\cite{1979Icar...40....1B} provide a simple expression for the critical value of $\beta$ for which cometary ejecta follow parabolic orbits.  However, particles tend to be ejected from comets, propelled by jets of gas, and their velocities therefore differ from that of the parent nucleus.  In this paper, we derive an analytic expression for the critical $\beta$ value for which cometary ejecta escape the solar system, taking this ejection velocity into account.  However, this analytic expression applies only to large particles governed by geometric optics.  Thus, we also extend our analysis to handle small particles and the non-geometric and non-ideal optics cases.  We present improved $\beta$ values for materials such as graphite, iron, and basalt.  We also correct the \cite{1995MNRAS.275..773J} model for the speed at which solid material is ejected by cometary nuclei, and extend it to handle small particles.

\section{Ejection velocity}

\cite{1951ApJ...113..464W} proposed a comet model in which meteoric materials are embedded in a nucleus comprised of frozen but volatile materials such as water and ammonia ices.  At small heliocentric distances, the volatile compounds sublimate from the sunlit side of the nucleus and expand outward in a coma.  These gases carry dust and meteoroids away from the nucleus; momentum is transferred to these solid particles through their absorption and re-emission of gas particles.  As the distance from the nucleus increases, the gas thins and momentum transfer ceases after having imparted to the particles an ejection velocity of $u_\infty$.

There are a number of variations on and parameterizations of the Whipple model 
\citep[examples include][]{1995MNRAS.275..773J,1998Icar..133...36B,2014ApJ...787L..35T,2014ApJ...792L..16K}.
These ejection velocity formulae generally depend on the size and density of the particle, heliocentric distance, and the size of the parent nucleus.  This paper presents a modification of the \cite{1995MNRAS.275..773J} model that corrects its coefficients and extends it to small particle sizes. A full derivation is presented in \ref{sec:uderiv}; here we present only those equations that are necessary for calculating the ejection speed.

It is first necessary to solve for the temperature, $T$, and saturation pressure, $p_\mathrm{sat}$, of the gas sublimating from the surface of the comet:
\begin{align}
    \frac{F}{4 r_h^2} &= \sigma T^4 + H(T) \, p_\mathrm{sat}(T) \sqrt{\frac{M}{2 \pi R T}} \label{eq:tsol} \\
    p_\mathrm{sat}(T) &= (3.56 \times 10^{12} \mbox{ Pa}) \, \cdot e^{-6141.667\mathrm{~K}/T}
\end{align}
where $F = 1361$~W~m$^{-2}$ is the solar irradiance at 1~au, $r_h$ is heliocentric distance in au, $\sigma$ is the Stefan-Boltzmann constant, $R$ is the gas constant, and $M$ is the molar mass of the gas. The quantity $H(T)$ is the heat of sublimation of water ice and can be taken from data tables or approximated as a constant (see Fig.~\ref{fig:Hg}). Eq.~\ref{eq:tsol} assumes that the incident solar radiation heats the entire comet nucleus evenly and that material is released from all directions. This is a simplification -- comets do not sublimate isotropically -- but jets of material have been seen to erupt from the dark sides of several comets, including 9P \citep{2013Icar..222..540F}, 103P \citep{2012DPS....4450601P}, and 67P \citep{2019A&A...630A..21R}. These jets sometimes occur hours after local sunset.

Once $T$ and $p_\mathrm{sat}$ have been determined, one can compute the speed, $v_1$, and density, $\rho_1$, of the gas near the comet after thermodynamic equilibrium has been re-established:
\begin{align}
    v_1(T) &= \sqrt{\pi R T / 2 M} \label{eq:v1} \\
    \rho_1(T) &= p_\mathrm{sat}(T) \cdot (M / \pi R T) \label{eq:p1}
\end{align}
Eqs.~\ref{eq:v1} and \ref{eq:p1} further assume that no gas particles re-condense onto the comet's surface.

Solid particles are assumed to be carried away from the nucleus by gas drag. The final speed of these particles relative to the comet, $u_\infty$, is:
\begin{align}
u_\infty &\simeq v_1(T) \left({0.4025 + 0.5139 \cdot \xi^{-1.054}}\right)^{-0.949} \label{eq:uinf} \\
\xi^2 &= \frac{A \Gamma}{2} m^{-1/3} \rho_m^{-2/3} \rho_1(T) \, R_c \label{eq:xi2}
\end{align}
where the dimensionless parameter $\xi$ encompasses all dependence on terms such as the particles' shape factor, $A = 1.209$; drag coefficient, $\Gamma = 2.89$; mass, $m$; bulk density, $\rho_m$; and the radius of the comet nucleus, $R_c$. Note that when $\xi$ is large, the first term in Eq.~\ref{eq:uinf} (i.e., the constant 0.4025) dominates; this corresponds to the limit in which the particles are small and move with the gas. When $\xi$ is small, the second term in  Eq.~\ref{eq:uinf} (i.e., $0.5139 \cdot \xi^{-1.054}$) dominates; this corresponds to the limit in which the particles are large and always move at speeds much lower than the gas speed.
 
Other authors have further simplified these equations by approximating $T(r_h)$ as a Taylor series about $r_h = 1$ and neglecting the small-particle component of Eq.~\ref{eq:uinf}, but we will use Eqs.~\ref{eq:tsol}--\ref{eq:xi2} in the form presented here.

\section{Escape velocity}
\label{sec:vesc}

Radiation pressure is a repulsive central force on small particles while gravity is an attractive central force. Both forces follow inverse square laws, and thus the ratio of radiation pressure to gravity provides us with a dimensionless parameter $\beta$, defined as
\begin{align}
\beta &= (3 L/16 \pi G M_\odot c)(Q_{pr}/\rho_m s) \nonumber \\
 &= 5.74 \times 10^{-4} \mbox{~kg m}^{-2} \times (Q_{pr}/\rho_m s) \label{eq:beta}
\end{align}
where $L = F \cdot 4 \pi \, \mathrm{au}^2$ is the solar luminosity, $G$ is the gravitational constant, $M_\odot$ is the Sun's mass, $c$ is the speed of light, $\rho_m$ is the bulk density of the affected particle, and $s$ is its radius \citep{1979Icar...40....1B}.  The final term, $Q_{pr}$, is the radiation pressure efficiency factor, which is equal to one for a perfectly absorbing particle that is significantly larger than the radiation wavelength (i.e., the ``geometric optics'' case).  

For a particle with non-zero $\beta$, the central potential is reduced by a factor $(1-\beta)$ and its escape speed is correspondingly reduced by a factor of $\sqrt{1-\beta}$.  The full expression for the escape speed at heliocentric distance $r = q$, where $q$ is the parent comet's perihelion distance, is:
\begin{align}
v_{esc}^2 &= G M_\odot (1-\beta) \frac{2}{q} \label{eq:vesc}
\end{align}
Particles are unbound from the Solar System when their total speed is greater than this escape speed.

This section presents our methods for calculating $Q_{pr}$ and thus $\beta$ for three cases: geometric optics, ideal materials, and non-ideal (or ``real") materials.

\subsection{Geometric optics}

In the geometric optics case, $Q_{pr} = 1$ at all wavelengths.  In this case, $\beta$ is extremely simple to calculate:
\begin{align}
\beta &= 5.74 \times 10^{-4} \mbox{~kg m}^{-2} \times (\rho_m s)^{-1} \label{eq:betageo}
\end{align}
This simple case is frequently invoked for dynamical studies of large meteoroids.

\subsection{Ideal material}

\begin{figure*} \centering
\includegraphics{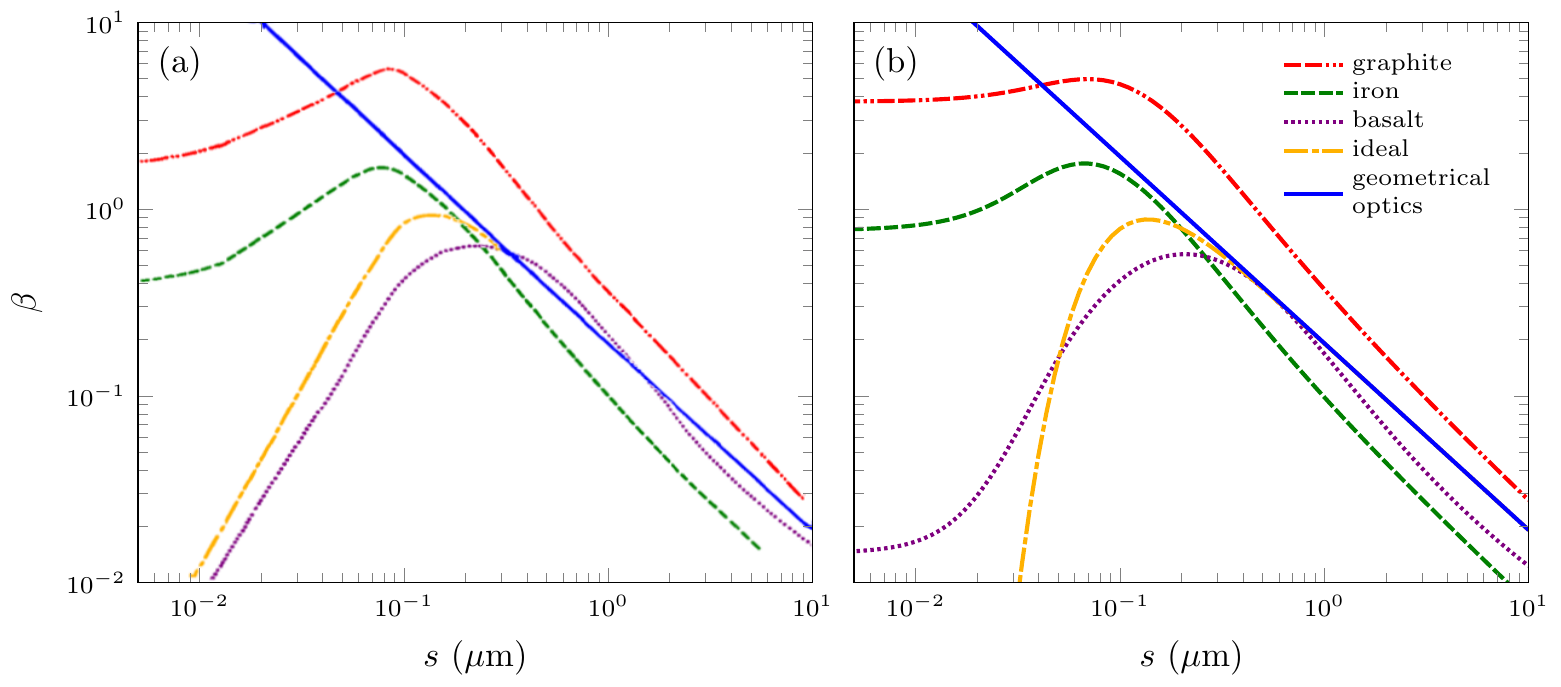}
\caption{Ratio of radiation pressure to gravity, $\beta$, as a function of particle size for several materials.  Panel (a) contains a colorized version of Fig.~7a from \cite{1979Icar...40....1B}.  Panel (b) presents our calculations for the same materials, using the same densities as \cite{1979Icar...40....1B}.}
\label{fig:bls}
\end{figure*}

When the particle size is comparable to the dominant wavelengths of the solar spectrum, $Q_{pr}$ deviates from 1 and the particle absorbs and scatters light less readily.  As a result, $\beta$ can begin to drop again for small particle sizes (see Fig.~\ref{fig:bls}).  Given the lack of detailed meteoroid optical information, \cite{1975P&SS...23..183Z} proposed a so-called ``ideal material" for which:
\begin{align}
Q_{pr}(\lambda) = \begin{cases}
  1 & \lambda < 2 \pi s \\
  0 & \lambda > 2 \pi s \label{eq:qideal}
\end{cases}
\end{align}
where $\lambda$ is wavelength.  Even this simple material is wavelength-dependent and thus Eq.~\ref{eq:qideal} must be convolved with the solar spectrum to obtain $Q_{pr}$: 
\begin{align}
Q_{pr} &= \frac{\int_0^\infty{Q_{pr}(\lambda') B_\lambda(\lambda',T) d \lambda'}}{
\int_0^\infty{B_\lambda(\lambda',T) d \lambda'}} \label{eq:qtot}
\end{align}
where $B_\lambda$ is the spectral radiance per unit wavelength.  For simplicity, we approximate the solar spectrum as a black body with a temperature of 5778 K.  This gives weighting factors similar to that shown in Fig.~18 of \cite{1975P&SS...23..183Z}.

Using Eq.~\ref{eq:qtot}, we can calculate $\beta$ as a function of particle size $s$ for this ``ideal material" and compare to \cite{1979Icar...40....1B}; Fig.~\ref{fig:bls} shows both solutions.  Here, we have used a density of 3000 kg m$^{-3}$ for the sake of comparing with the earlier work.  While both are equivalent to the geometrical optics case at large sizes, our solution deviates from that of \cite{1979Icar...40....1B} for small particles.  The reason for this discrepancy is unclear.

\subsubsection{Real materials}

Real materials can behave very differently from an ideal material, and both the refractive index and the extinction coefficient (i.e., the real and imaginary components of the complex index of refraction) can vary with wavelength.  We calculate $Q_{pr}$ as a function of wavelength and particle size using code provided by \cite{Navarro2013}, which is itself a translation to Python of M\"{a}tzler's MATLAB package \citep{Maetzlerv1}.  We have modified the code slightly to reduce underflow errors.

The \cite{Navarro2013} code computes absorption and scattering terms ($Q_{abs}$ and $Q_{sca}$) as well as the asymmetry parameter $\langle \cos \phi \rangle$ for a given $x = 2 \pi s / \lambda$ and complex index of refraction, $\vec{n}(\lambda)$.  The radiation pressure efficiency factor is obtained from these terms as follows:
\begin{align}
Q_{pr} &= Q_{abs} - Q_{sca} \langle \cos \phi \rangle \label{eq:qtot2}
\end{align}
Once again, we convolve Eq.~\ref{eq:qtot2} with a black-body solar spectrum to obtain the total effective $Q_{pr}$.

As mentioned, the \cite{Navarro2013} code requires a complex index of refraction for the material in question.  \cite{1985umo..rept.....Q} provides these indices for iron and graphite pellets as a function of wavelength in the range 0.21-55.5~$\mu$m.  Figure \ref{fig:bls} shows the resulting $\beta$ values.  We have assumed a density of 2150~kg~m$^{-3}$ for graphite and 7870~kg~m$^{-3}$ for iron.  Our $\beta$ calculations for both materials have the same general shape as those of \cite{1979Icar...40....1B}, although we obtained higher values of $\beta$ at small sizes. 

\begin{figure} \centering
\includegraphics{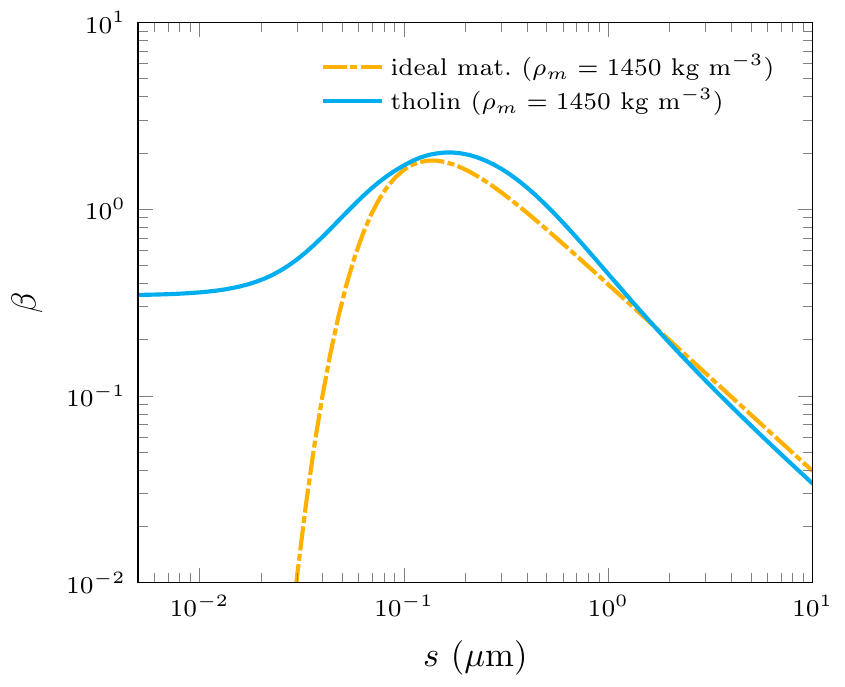}
\caption{Ratio of radiation pressure to gravity, $\beta$, as a function of particle size for two materials of the same density.  We compare an ideal material with a density of 1450~kg~m$^{-3}$ to a tholin of the same bulk density.  Values of $\beta$ for the two materials are similar at large sizes but deviate significantly at small sizes.}
\label{fig:btholin}
\end{figure}

We have also computed $\beta$ for an additional material, tholins \citep{1979Natur.277..102S}, which are thought to occur on the surface of many icy bodies in the solar system, including comets. \cite{1984Icar...60..127K} provide values for the complex refractive index of tholins over a wide range of wavelengths.  We assume a density of 1450~kg~m$^{-3}$ for the tholin grains \citep{1988Icar...76..100L}, which also falls within the range of acceptable meteoroid densities \citep{2011A&A...530A.113K,stx2175}.  In Fig.~\ref{fig:btholin}, we compare these tholin $\beta$ values with those of an ideal material of the same density.  We obtain values that are very similar to those of \cite{1988Icar...76..100L}.  At micron-or-larger sizes, the tholins and the ideal material have very similar $\beta$ values, but the behavior at small sizes is significantly different.  Thus, while sub-micron ideal material grains will be little affected by radiation pressure, sub-micron tholin grains will experience a reduced escape velocity compared to their parent comet.

\section{Unbound mass ranges}

Many meteor shower parent bodies have high eccentricities and their speeds at pericenter can be quite close to escape velocity.  The velocity of a comet (or any orbiting object) at perihelion is:
\begin{align}
v_p^2 &= G M_\odot \left({\frac{2}{q}-\frac{1}{a}}\right) \label{eq:vp}
\end{align}
where $q$ is perihelion distance and $a$ is semi-major axis.

This speed is equal to the local escape speed (Eq.~\ref{eq:vesc}) when $\beta = \beta_0 = q/2 a = (1-e)/2$, where $e$ is orbital eccentricity, as given by \cite{1979Icar...40....1B}.  Particles with this value of $\beta$ or larger are unbound from the Sun if their speed matches that of the comet nucleus. In this section, we also consider cases in which particles are ejected from the comet with some additional velocity.  Depending on the direction of ejection, this effect may either raise or lower the mass threshold at which particles become unbound.

We combine our calculations of the particle ejection speed and escape velocity in order to compute the particle mass ranges that are excluded due to radiation pressure. For large particles that obey geometric optics, an analytic solution exists; we present this solution in \ref{sec:large} for interested readers. In all other cases, it is necessary to compute the range of excluded masses numerically. To do so, we solve for ejection velocity (using Eqs.~\ref{eq:uinf} and \ref{eq:xi2}) and escape velocity (using Eqs.~\ref{eq:beta} and \ref{eq:vesc}) as a function of particle size.  By comparing the two velocities, we can determine the size ranges within which all or some ejected particles will be on unbound orbits and thus lost from the solar system.  

\begin{figure} \centering
\includegraphics{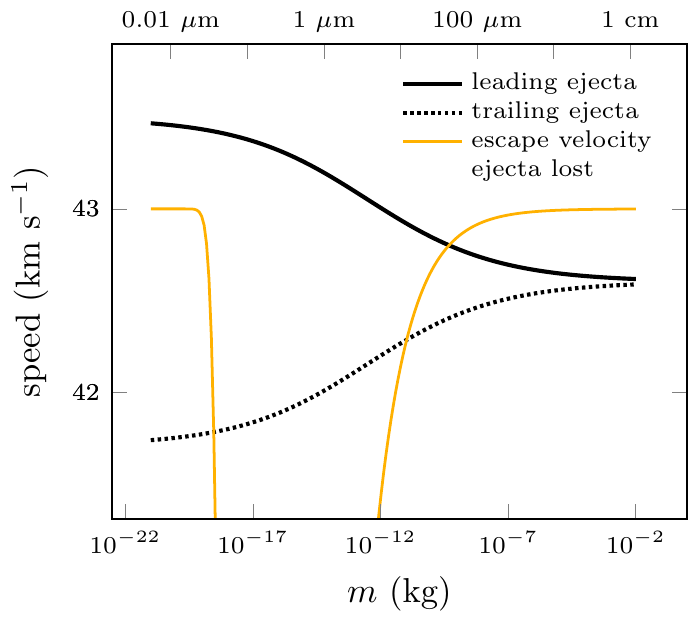}
\vspace{0.1in} \\
\includegraphics{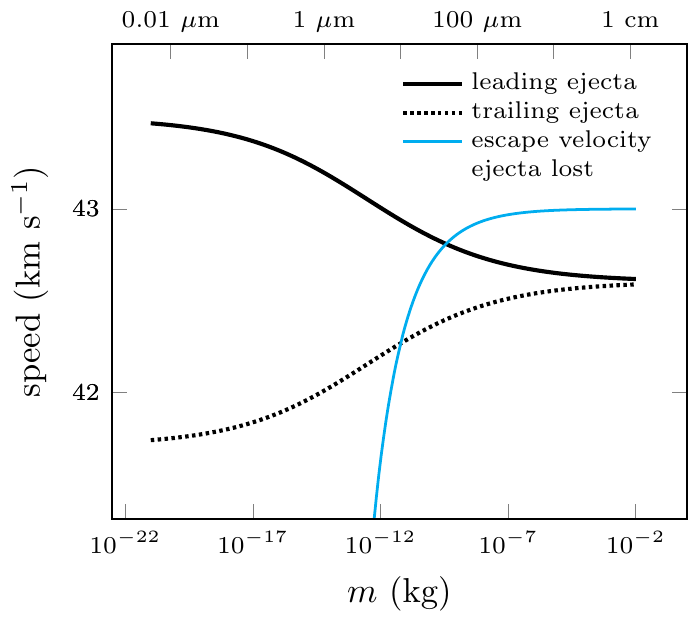}
\caption{This plot compares the initial velocity of Perseid meteoroids and dust with the velocity of escape from the solar system as a function of particle mass or radius, assuming that all particles are composed of ideal material (top) or tholins (bottom) and have a density of 1450~kg~m$^{-3}$. The black lines represent the total velocity of particles ejected in (leading ejecta) and opposite to (trailing ejecta) the direction of the parent comet's motion at perihelion.  The hatched regions indicate material whose velocity exceeds escape velocity and is thus lost due to radiation pressure.}
\label{fig:peri}
\end{figure}

Figure~\ref{fig:peri} compares these speeds for the Perseids, which originate from comet 109P/Swift-Tuttle.  The top panel of Fig.~\ref{fig:peri} demonstrates that for ejecta composed of ideal material, particles smaller than about $4 \times 10^{-19}$~kg are removed partly or wholly from the stream. While trailing ejecta smaller than $2 \times 10^{-19}$~kg are retained, the large size of Swift-Tuttle produces ejection velocities that are large enough to remove all leading ejecta less massive than $4 \times 10^{-10}$ kg. The lower panel compares the same Perseid ejection velocities with the escape velocity corresponding to tholins.  In this case, all material below $7 \times 10^{-12}$ kg is removed from the stream.  This is due to tholin grains having significantly larger values of $\beta$ for small particles (see Fig.~\ref{fig:btholin}).

\begin{figure*} \centering
    \includegraphics{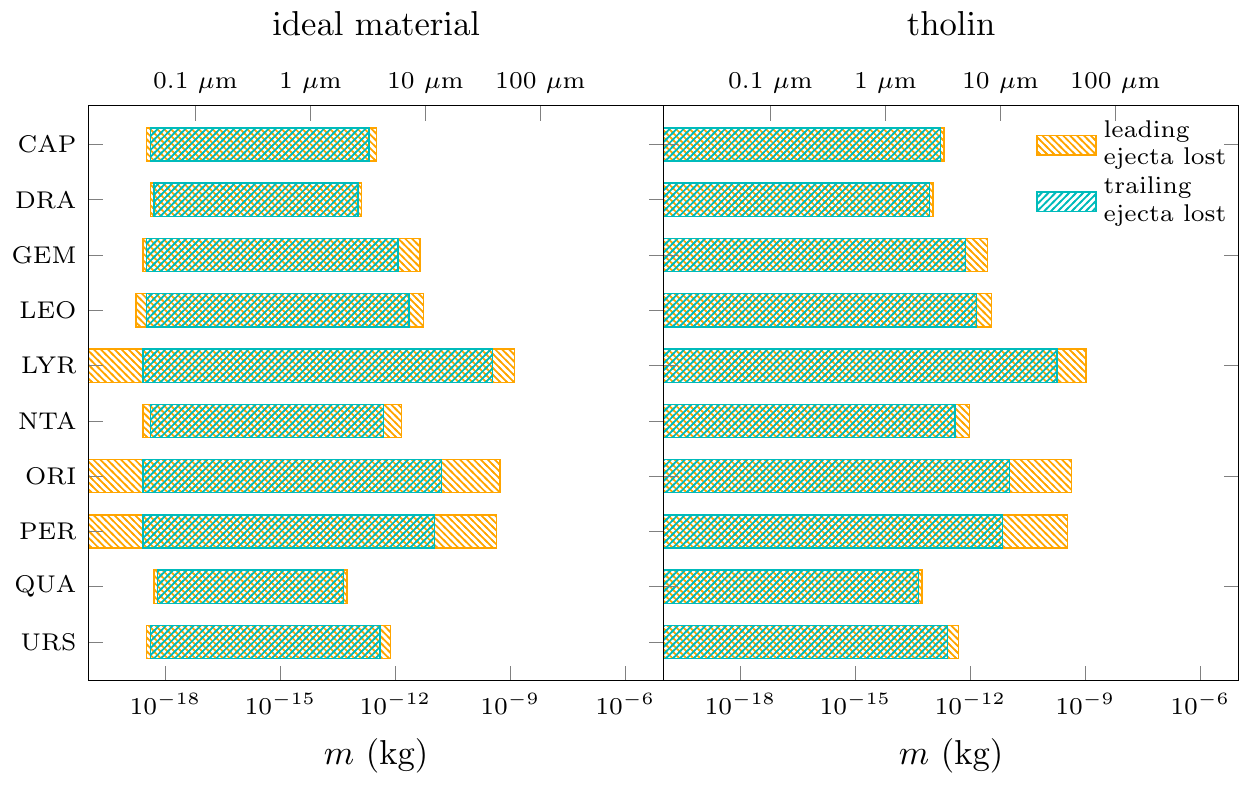}
    \caption{The above bar charts indicate the mass (or radius) ranges at which particles cannot form a part of a given meteor shower. Ten showers are considered here: the alpha Capricornids (CAP), the Draconids (DRA), the Geminids (GEM), the Leonids (LEO), the April Lyrids (LYR), the Northern Taurids (NTA; calculation also applies to the Southern Taurids), the Orionids (ORI), the Perseids (PER), the Quadrantids (QUA), and the Ursids (URS). The excluded mass range for each shower is calculated assuming ideal optics or that the particles have the optical properties of tholins. A density of 1450~kg~m$^{-3}$ was used in both cases.}
    \label{fig:bar}
\end{figure*}

Figure \ref{fig:bar} summarizes our results for 10 major meteor showers with known parent bodies.  For each shower, we present the mass interval (or intervals) within which meteors are not bound to the solar system.   We present the results for both an ideal material and a tholin.  Both cases assume a density of 1450~kg~m$^{-3}$, which corresponds to the results shown in Figures~\ref{fig:btholin}-\ref{fig:peri}.  The upper mass limit at which streams begin to lose particles is comparable to those calculated by \cite{1970JGR....75.3468D} assuming geometric optics and neglecting ejection velocity.

The extent of the excluded mass ranges in Fig.~\ref{fig:bar} depends on both the parent comet properties and the particle properties. For instance, the parent body of the Lyrids (LYR) is comet C/1861 G1 (Thatcher); this comet has an eccentricity of 0.983, the highest of all parent bodies considered here, and is thus moving very close to escape velocity at perihelion. As a result, this stream begins to lose leading particles around 10$^{-9}$~kg. The optical properties of the ejecta also affect these limits, particularly for very small dust particles. Some fraction of the smallest dust particles in our plots remain in the stream if they have ideal optical properties, but are not retained if they resemble tholins (or graphite or iron). The density of the particles can also shift these ranges by a factor of a few (i.e., 2-4). For instance, in the left panel of Fig.~\ref{fig:bar}, Geminids with masses between $3.29 \times 10^{-19}$~kg and $1.18 \times 10^{-12}$~kg are excluded from the stream entirely. If we instead use a density of 3000~kg~m$^{-3}$ \citep[which is more in line with current estimates of Geminid densities;][]{2010IAUS..263..218B,2019P&SS..173...42N}, this range shrinks to [$9.83 \times 10^{-19}$~kg, $3.18 \times 10^{-13}$~kg]. Because these shifts are small compared to the range depicted in Fig.~\ref{fig:bar}, and because the overall behavior is unchanged, we have opted not to depict them.

\section{Limitations and implications}

This study has a number of limitations that we would like to highlight here. First, we neglect the influence of the Lorentz force on small particles \citep{1980Icar...43..203C}. This effect, in which particles accumulate charge and interact with the Sun's magnetic field, can produce significant changes in the orbits of small particles. \cite{1993mtpb.conf..381I} find that for basalt-like particles, Lorentz forces can be more significant than radiation pressure for particles smaller than $1~\mu$m during solar maximum. The effect of the Lorentz force varies over the solar cycle and also depends on the photoelectric yield of the particle material; overall, it induces a random walk on the orbital elements of small particles \citep{1980Icar...43..203C}. The Lorentz force may result in the loss of additional small particles, but, due to its variability, we do not believe it can produce long-term retention of particles whose orbital velocity exceeds the local escape velocity at perihelion.

Second, we treat stream membership very simply: a particle ejected from a meteor shower's parent body is considered to be part of the stream unless its eccentricity is greater than one. We make no attempt to define the point at which a bound particle becomes part of the sporadic complex. Similarly, we have not attempted to define when a particle does not contribute to meteor showers \emph{seen at Earth}. Those particles that remain bound to the solar system will still follow orbits that differ from those of their parent body due to the effects of ejection and radiation pressure. Particles that form part of inclined streams may intersect the ecliptic plane at a range of heliocentric distances; this could cause certain particle ranges to be preferentially seen or not seen at the Earth; \cite{1999PhDT.........7B} demonstrates how radiation pressure and ejection velocity correspond to a spread in nodal radius.

\begin{figure} \centering
\includegraphics[width=0.9\columnwidth]{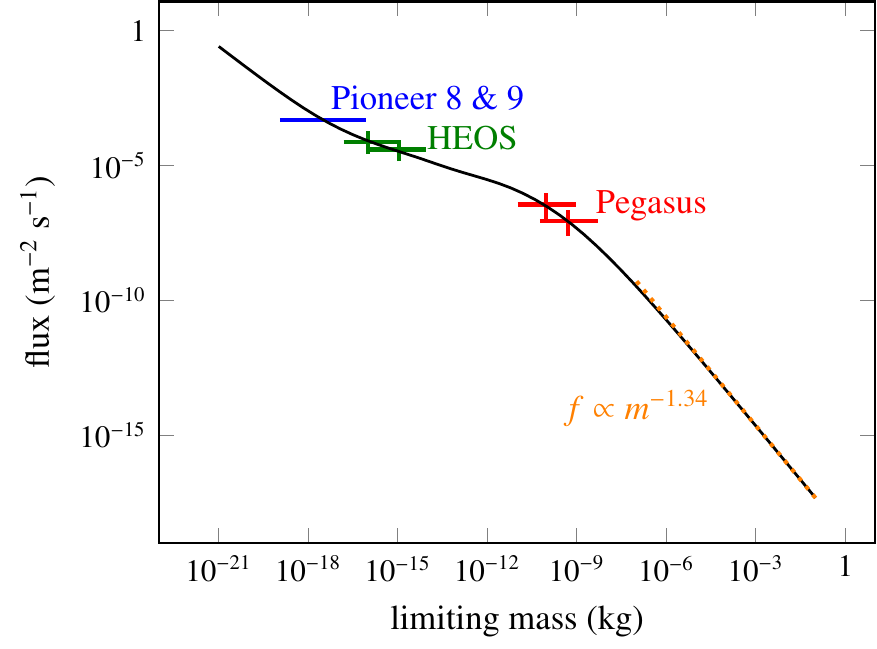}
\caption{The flux model of \cite{1985Icar...62..244G} is shown here as a black curve. We also present the key data points used to constrain that model; this figure corresponds to Fig.~1 of \cite{1985Icar...62..244G}. We have included the sporadic mass index of \cite{1958ApJ...128..727H} as a dotted orange line.}
\label{fig:grun}
\end{figure}

The inclusion or exclusion of meteoroids of a certain size from meteoroid streams has ramifications for our understanding of the meteoroid environment.  Many studies use Equation A3 of \cite{1985Icar...62..244G} to describe the interplanetary meteoroid flux at 1~au as a function of particle mass; this equation is presented in Fig.~\ref{fig:grun}.  \citeauthor{1985Icar...62..244G} constrained this flux curve using Pioneer, HEOS, and Pegasus data at the small end, and tied the slope at large particle sizes to the mass index of 1.34 measured by \cite{1958ApJ...128..727H} and quoted by \cite{1967SAOSR.239....1W}.  This mass index of 1.34 was derived using 300 sporadic meteors \citep[see the bottom of page 729 of][]{1958ApJ...128..727H}. 

Figure~\ref{fig:bar} indicates that particles ranging in mass from approximately 10$^{-18}$ to 10$^{-13}$~kg are excluded from all ten streams by radiation pressure. This mass range encompasses the limiting masses measured by {\sl in situ} impact experiments such as Pioneer 8 and 9 and HEOS. Furthermore, several showers -- the Lyrids, Orionids, and Perseids -- lose a portion of ejected particles smaller than 10$^{-9}$~kg, possibly reducing the contribution of meteoroid streams to the Pegasus data. In fact, an unpublished reanalysis of Pegasus data revealed no shower signatures (Blaauw, private communication). Thus, the \cite{1985Icar...62..244G} flux model most likely describes only sporadic meteoroids and not the meteoroid environment as a whole.

This in turn has implications for meteor shower forecasting. The NASA Meteoroid Environment Office issues annual meteor shower forecasts to facilitate spacecraft risk assessments. These forecasts provide an enhancement factor that represents the percentage by which the total flux, sporadic plus shower, lies above or below the \cite{1985Icar...62..244G} flux. Historically, the forecasting algorithm assumes that showers contribute to the \cite{1985Icar...62..244G} flux, and thus times of low shower activity correspond to negative enhancement factors \citep{moorheadECSD}. However, this paper indicates that it would be more accurate to consider any shower activity as an enhancement over \cite{1985Icar...62..244G}; we recently updated our shower forecasting algorithms to adopt this approach \citep{2019JSpRo..56.1531M}.

\section{Conclusions}

In this paper we have considered the processes by which particles may be removed from a meteoroid stream in less than one orbit by radiation pressure. We include the velocity at which a comet nucleus releases ejecta in order to determine the mass range within some or all particles are lost from the stream.

We have re-derived the \cite{1995MNRAS.275..773J} meteoroid ejection velocity equation for the case of spherically symmetric ejection, extending it to include small particles. We have also computed the radiation pressure efficiency factor, $Q_{pr}$, for a selection of meteoroid material analogs. This factor is critical for obtaining the relative importance of radiation pressure and gravity for small particles, $\beta$. We have reproduced some of the features of $\beta$ seen in the work of \cite{1979Icar...40....1B}, and have extended the set of materials considered to include tholins.

We find that for the ten major meteoroid streams we consider, particles between 10$^{-18}$ and 10$^{-13}$~kg are lost due to radiation pressure. In some streams, some particles as large as 10$^{-9}$~kg are lost due to the combined effects of radiation pressure and ejection velocity. Very small particles ($<10^{-18}$~kg) may be initially retained if they behave like so-called ``ideal'' optical material, but will likely be subsequently lost due to the Lorentz force.

These boundaries may be blurred or expanded by the subsequent orbital and material evolution of meteoroids and dust particles. Particle orbits will be modified over time by the gravitational influence of the planets, Poynting-Robertson drag, solar wind pressure, and, for small particles, Lorentz forces, but the ability of radiation pressure to remove particles in less than one orbit is difficult to overcome through any of these effects. Over longer timescales, however, we hypothesize that collisions may be able to backfill the excluded mass ranges. We thus conclude that meteoroid showers are unlikely to contribute to the data used to construct the \cite{1985Icar...62..244G} meteoroid flux curve, making that study a sporadic-only model.

It may be possible to test our predictions using meteor radar shower surveys. For instance, the Middle Atmosphere ALOMAR Radar System (MAARSY) detected Lyrid meteors just above our predicted threshold; the mean ($\log_{10}$) dynamical mass detected was -8.1, with a standard deviation of 1.1. A radar with an order-of-magnitude lower mass threshold may be able to test whether the size distribution ``breaks'' near 10$^{-9}$~kg as predicted.

\section*{Acknowledgement}

The author would like to thank Jim Jones for helping them get to the bottom of a number of small discrepancies between this work and \cite{1995MNRAS.275..773J}.

\bibliographystyle{elsarticle-harv}
\bibliography{local}

\begin{thebibliography}{44}
\expandafter\ifx\csname natexlab\endcsname\relax\def\natexlab#1{#1}\fi
\providecommand{\url}[1]{\texttt{#1}}
\providecommand{\href}[2]{#2}
\providecommand{\path}[1]{#1}
\providecommand{\DOIprefix}{doi:}
\providecommand{\ArXivprefix}{arXiv:}
\providecommand{\URLprefix}{URL: }
\providecommand{\Pubmedprefix}{pmid:}
\providecommand{\doi}[1]{\href{http://dx.doi.org/#1}{\path{#1}}}
\providecommand{\Pubmed}[1]{\href{pmid:#1}{\path{#1}}}
\providecommand{\bibinfo}[2]{#2}
\ifx\xfnm\relax \def\xfnm[#1]{\unskip,\space#1}\fi
\bibitem[{{Benkhoff} and {Huebner}(1995)}]{1995Icar..114..348B}
\bibinfo{author}{{Benkhoff}, J.}, \bibinfo{author}{{Huebner}, W.F.},
  \bibinfo{year}{1995}.
\newblock \bibinfo{title}{{Influence of the Vapor Flux on Temperature, Density,
  and Abundance Distributions in a Multicomponent, Porous, Icy Body}}.
\newblock \bibinfo{journal}{Icarus} \bibinfo{volume}{114},
  \bibinfo{pages}{348--354}.
\newblock \DOIprefix\doi{10.1006/icar.1995.1067}.
\bibitem[{{Borovi{\v{c}}ka} et~al.(2010){Borovi{\v{c}}ka}, {Koten},
  {Spurn{\'y}}, {{\v{C}}apek}, {Shrben{\'y}} and
  {{\v{S}}tork}}]{2010IAUS..263..218B}
\bibinfo{author}{{Borovi{\v{c}}ka}, J.}, \bibinfo{author}{{Koten}, P.},
  \bibinfo{author}{{Spurn{\'y}}, P.}, \bibinfo{author}{{{\v{C}}apek}, D.},
  \bibinfo{author}{{Shrben{\'y}}, L.}, \bibinfo{author}{{{\v{S}}tork}, R.},
  \bibinfo{year}{2010}.
\newblock \bibinfo{title}{{Material properties of transition objects 3200
  Phaethon and 2003 EH$_{1}$}}, in: \bibinfo{editor}{{Fernandez}, J.A.},
  \bibinfo{editor}{{Lazzaro}, D.}, \bibinfo{editor}{{Prialnik}, D.},
  \bibinfo{editor}{{Schulz}, R.} (Eds.), \bibinfo{booktitle}{Icy Bodies of the
  Solar System}, pp. \bibinfo{pages}{218--222}.
\newblock \DOIprefix\doi{10.1017/S174392131000178X}.
\bibitem[{{Brandt} and {Chapman}(2004)}]{2004inco.book.....B}
\bibinfo{author}{{Brandt}, J.C.}, \bibinfo{author}{{Chapman}, R.D.},
  \bibinfo{year}{2004}.
\newblock \bibinfo{title}{{Introduction to Comets}}.
\bibitem[{{Brown} and {Jones}(1998)}]{1998Icar..133...36B}
\bibinfo{author}{{Brown}, P.}, \bibinfo{author}{{Jones}, J.},
  \bibinfo{year}{1998}.
\newblock \bibinfo{title}{{Simulation of the Formation and Evolution of the
  Perseid Meteoroid Stream}}.
\newblock \bibinfo{journal}{Icarus} \bibinfo{volume}{133},
  \bibinfo{pages}{36--68}.
\newblock \DOIprefix\doi{10.1006/icar.1998.5920}.
\bibitem[{{Brown}(1999)}]{1999PhDT.........7B}
\bibinfo{author}{{Brown}, P.G.}, \bibinfo{year}{1999}.
\newblock \bibinfo{title}{{Evolution of two periodic meteoroid streams: The
  Perseids and Leonids}}.
\newblock Ph.D. thesis. The University of Western Ontario (Canada).
\bibitem[{{Burns} et~al.(1979){Burns}, {Lamy} and
  {Soter}}]{1979Icar...40....1B}
\bibinfo{author}{{Burns}, J.A.}, \bibinfo{author}{{Lamy}, P.L.},
  \bibinfo{author}{{Soter}, S.}, \bibinfo{year}{1979}.
\newblock \bibinfo{title}{{Radiation forces on small particles in the solar
  system}}.
\newblock \bibinfo{journal}{Icarus} \bibinfo{volume}{40},
  \bibinfo{pages}{1--48}.
\newblock \DOIprefix\doi{10.1016/0019-1035(79)90050-2}.
\bibitem[{{Consolmagno}(1980)}]{1980Icar...43..203C}
\bibinfo{author}{{Consolmagno}, G.J.}, \bibinfo{year}{1980}.
\newblock \bibinfo{title}{{Influence of the interplanetary magnetic field on
  cometary and primordial dust orbits: Applications of Lorentz scattering}}.
\newblock \bibinfo{journal}{Icarus} \bibinfo{volume}{43},
  \bibinfo{pages}{203--214}.
\newblock \DOIprefix\doi{10.1016/0019-1035(80)90121-9}.
\bibitem[{{Davidsson} and {Skorov}(2004)}]{2004Icar..168..163D}
\bibinfo{author}{{Davidsson}, B.J.R.}, \bibinfo{author}{{Skorov}, Y.V.},
  \bibinfo{year}{2004}.
\newblock \bibinfo{title}{{A practical tool for simulating the presence of gas
  comae in thermophysical modeling of cometary nuclei}}.
\newblock \bibinfo{journal}{Icarus} \bibinfo{volume}{168},
  \bibinfo{pages}{163--185}.
\newblock \DOIprefix\doi{10.1016/j.icarus.2003.11.002}.
\bibitem[{{Dohnanyi}(1970)}]{1970JGR....75.3468D}
\bibinfo{author}{{Dohnanyi}, J.S.}, \bibinfo{year}{1970}.
\newblock \bibinfo{title}{{On the origin and distribution of meteoroids.}}
\newblock \bibinfo{journal}{J.~Geophys.~Res.} \bibinfo{volume}{75},
  \bibinfo{pages}{3468--3493}.
\newblock \DOIprefix\doi{10.1029/JB075i017p03468}.
\bibitem[{{Fanale} and {Salvail}(1984)}]{1984Icar...60..476F}
\bibinfo{author}{{Fanale}, F.P.}, \bibinfo{author}{{Salvail}, J.R.},
  \bibinfo{year}{1984}.
\newblock \bibinfo{title}{{An idealized short-period comet model: Surface
  insolation, H$_{2}$O flux, dust flux, and mantle evolution}}.
\newblock \bibinfo{journal}{Icarus} \bibinfo{volume}{60},
  \bibinfo{pages}{476--511}.
\newblock \DOIprefix\doi{10.1016/0019-1035(84)90157-X}.
\bibitem[{{Farnham} et~al.(2013){Farnham}, {Bodewits}, {Li}, {Veverka},
  {Thomas} and {Belton}}]{2013Icar..222..540F}
\bibinfo{author}{{Farnham}, T.L.}, \bibinfo{author}{{Bodewits}, D.},
  \bibinfo{author}{{Li}, J.Y.}, \bibinfo{author}{{Veverka}, J.},
  \bibinfo{author}{{Thomas}, P.}, \bibinfo{author}{{Belton}, M.J.S.},
  \bibinfo{year}{2013}.
\newblock \bibinfo{title}{{Connections between the jet activity and surface
  features on Comet 9P/Tempel 1}}.
\newblock \bibinfo{journal}{Icarus} \bibinfo{volume}{222},
  \bibinfo{pages}{540--549}.
\newblock \DOIprefix\doi{10.1016/j.icarus.2012.06.019}.
\bibitem[{{Groussin} et~al.(2007){Groussin}, {A'Hearn}, {Li}, {Thomas},
  {Sunshine}, {Lisse}, {Meech}, {Farnham}, {Feaga} and
  {Delamere}}]{2007Icar..187...16G}
\bibinfo{author}{{Groussin}, O.}, \bibinfo{author}{{A'Hearn}, M.F.},
  \bibinfo{author}{{Li}, J.Y.}, \bibinfo{author}{{Thomas}, P.C.},
  \bibinfo{author}{{Sunshine}, J.M.}, \bibinfo{author}{{Lisse}, C.M.},
  \bibinfo{author}{{Meech}, K.J.}, \bibinfo{author}{{Farnham}, T.L.},
  \bibinfo{author}{{Feaga}, L.M.}, \bibinfo{author}{{Delamere}, W.A.},
  \bibinfo{year}{2007}.
\newblock \bibinfo{title}{{Surface temperature of the nucleus of Comet
  9P/Tempel 1}}.
\newblock \bibinfo{journal}{Icarus} \bibinfo{volume}{187},
  \bibinfo{pages}{16--25}.
\newblock \DOIprefix\doi{10.1016/j.icarus.2006.08.030}.
\bibitem[{{Gr\"{u}n} et~al.(1985){Gr\"{u}n}, {Zook}, {Fechtig} and
  {Giese}}]{1985Icar...62..244G}
\bibinfo{author}{{Gr\"{u}n}, E.}, \bibinfo{author}{{Zook}, H.A.},
  \bibinfo{author}{{Fechtig}, H.}, \bibinfo{author}{{Giese}, R.H.},
  \bibinfo{year}{1985}.
\newblock \bibinfo{title}{{Collisional balance of the meteoritic complex}}.
\newblock \bibinfo{journal}{Icarus} \bibinfo{volume}{62},
  \bibinfo{pages}{244--272}.
\newblock \DOIprefix\doi{10.1016/0019-1035(85)90121-6}.
\bibitem[{{Harwit}(1963)}]{1963JGR....68.2171H}
\bibinfo{author}{{Harwit}, M.}, \bibinfo{year}{1963}.
\newblock \bibinfo{title}{{Origins of the Zodiacal Dust Cloud}}.
\newblock \bibinfo{journal}{Journal of Geophysical Research}
  \bibinfo{volume}{68}, \bibinfo{pages}{2171--2180}.
\newblock \DOIprefix\doi{10.1029/JZ068i008p02171}.
\bibitem[{{Hawkins} and {Upton}(1958)}]{1958ApJ...128..727H}
\bibinfo{author}{{Hawkins}, G.S.}, \bibinfo{author}{{Upton}, E.K.L.},
  \bibinfo{year}{1958}.
\newblock \bibinfo{title}{{The Influx Rate of Meteors in the Earth's
  Atmosphere.}}
\newblock \bibinfo{journal}{Astrophys.~J.} \bibinfo{volume}{128},
  \bibinfo{pages}{727}.
\newblock \DOIprefix\doi{10.1086/146585}.
\bibitem[{{Huebner} and {Markiewicz}(2000)}]{2000Icar..148..594H}
\bibinfo{author}{{Huebner}, W.F.}, \bibinfo{author}{{Markiewicz}, W.J.},
  \bibinfo{year}{2000}.
\newblock \bibinfo{title}{{NOTE: The Temperature and Bulk Flow Speed of a Gas
  Effusing or Evaporating from a Surface into a Void after Reestablishment of
  Collisional Equilibrium}}.
\newblock \bibinfo{journal}{Icarus} \bibinfo{volume}{148},
  \bibinfo{pages}{594--596}.
\newblock \DOIprefix\doi{10.1006/icar.2000.6522}.
\bibitem[{Ismail et~al.(2015)Ismail, Khulbe and Matsuura}]{Ismail2015}
\bibinfo{author}{Ismail, A.F.}, \bibinfo{author}{Khulbe, K.C.},
  \bibinfo{author}{Matsuura, T.}, \bibinfo{year}{2015}.
\newblock \bibinfo{title}{Gas Separation Membranes}.
\newblock \bibinfo{publisher}{Springer International Publishing}.
\newblock \URLprefix \url{https://doi.org/10.1007/978-3-319-01095-3},
  \DOIprefix\doi{10.1007/978-3-319-01095-3}.
\bibitem[{{Isobe} and {Sateesh-Kumar}(1993)}]{1993mtpb.conf..381I}
\bibinfo{author}{{Isobe}, S.}, \bibinfo{author}{{Sateesh-Kumar}, A.},
  \bibinfo{year}{1993}.
\newblock \bibinfo{title}{{An effect of Lorentz force on interplanetary dust}},
  in: \bibinfo{editor}{{Stohl}, J.}, \bibinfo{editor}{{Williams}, I.P.} (Eds.),
  \bibinfo{booktitle}{Meteoroids and their Parent Bodies}, p.
  \bibinfo{pages}{381}.
\bibitem[{Jackson(1998)}]{JacksonEM}
\bibinfo{author}{Jackson, J.D.}, \bibinfo{year}{1998}.
\newblock \bibinfo{title}{Classical Electrodynamics Third Edition}.
\newblock \bibinfo{publisher}{Wiley}.
\newblock \URLprefix \url{https://www.xarg.org/ref/a/047130932X/}.
\bibitem[{{Jones}(1995)}]{1995MNRAS.275..773J}
\bibinfo{author}{{Jones}, J.}, \bibinfo{year}{1995}.
\newblock \bibinfo{title}{{The ejection of meteoroids from comets}}.
\newblock \bibinfo{journal}{MNRAS} \bibinfo{volume}{275},
  \bibinfo{pages}{773--780}.
\newblock \DOIprefix\doi{10.1093/mnras/275.3.773}.
\bibitem[{{Kelley} et~al.(2014){Kelley}, {Farnham}, {Bodewits}, {Tricarico} and
  {Farnocchia}}]{2014ApJ...792L..16K}
\bibinfo{author}{{Kelley}, M.S.P.}, \bibinfo{author}{{Farnham}, T.L.},
  \bibinfo{author}{{Bodewits}, D.}, \bibinfo{author}{{Tricarico}, P.},
  \bibinfo{author}{{Farnocchia}, D.}, \bibinfo{year}{2014}.
\newblock \bibinfo{title}{{A Study of Dust and Gas at Mars from Comet C/2013 A1
  (Siding Spring)}}.
\newblock \bibinfo{journal}{Astrophys.~J.l} \bibinfo{volume}{792},
  \bibinfo{pages}{L16}.
\newblock \DOIprefix\doi{10.1088/2041-8205/792/1/L16},
  \href{http://arxiv.org/abs/1408.2792}{{\tt arXiv:1408.2792}}.
\bibitem[{{Khare} et~al.(1984){Khare}, {Sagan}, {Arakawa}, {Suits}, {Callcott}
  and {Williams}}]{1984Icar...60..127K}
\bibinfo{author}{{Khare}, B.N.}, \bibinfo{author}{{Sagan}, C.},
  \bibinfo{author}{{Arakawa}, E.T.}, \bibinfo{author}{{Suits}, F.},
  \bibinfo{author}{{Callcott}, T.A.}, \bibinfo{author}{{Williams}, M.W.},
  \bibinfo{year}{1984}.
\newblock \bibinfo{title}{{Optical constants of organic tholins produced in a
  simulated Titanian atmosphere - From soft X-ray to microwave frequencies}}.
\newblock \bibinfo{journal}{Icarus} \bibinfo{volume}{60},
  \bibinfo{pages}{127--137}.
\newblock \DOIprefix\doi{10.1016/0019-1035(84)90142-8}.
\bibitem[{{Kikwaya} et~al.(2011){Kikwaya}, {Campbell-Brown} and
  {Brown}}]{2011A&A...530A.113K}
\bibinfo{author}{{Kikwaya}, J.B.}, \bibinfo{author}{{Campbell-Brown}, M.},
  \bibinfo{author}{{Brown}, P.G.}, \bibinfo{year}{2011}.
\newblock \bibinfo{title}{{Bulk density of small meteoroids}}.
\newblock \bibinfo{journal}{Astron.~{\&}~Astrophys.} \bibinfo{volume}{530},
  \bibinfo{pages}{A113}.
\newblock \DOIprefix\doi{10.1051/0004-6361/201116431}.
\bibitem[{{Lamy} and {Perrin}(1988)}]{1988Icar...76..100L}
\bibinfo{author}{{Lamy}, P.L.}, \bibinfo{author}{{Perrin}, J.M.},
  \bibinfo{year}{1988}.
\newblock \bibinfo{title}{{Optical properties of organic grains - Implications
  for interplanetary and cometary dust}}.
\newblock \bibinfo{journal}{Icarus} \bibinfo{volume}{76},
  \bibinfo{pages}{100--109}.
\newblock \DOIprefix\doi{10.1016/0019-1035(88)90142-X}.
\bibitem[{{Ma} et~al.(2002){Ma}, {Williams} and {Chen}}]{2002MNRAS.337.1081M}
\bibinfo{author}{{Ma}, Y.}, \bibinfo{author}{{Williams}, I.P.},
  \bibinfo{author}{{Chen}, W.}, \bibinfo{year}{2002}.
\newblock \bibinfo{title}{{On the ejection velocity of meteoroids from
  comets}}.
\newblock \bibinfo{journal}{MNRAS} \bibinfo{volume}{337},
  \bibinfo{pages}{1081--1086}.
\newblock \DOIprefix\doi{10.1046/j.1365-8711.2002.05996.x}.
\bibitem[{{M\"{a}tzler}(2002)}]{Maetzlerv1}
\bibinfo{author}{{M\"{a}tzler}, C.}, \bibinfo{year}{2002}.
\newblock \bibinfo{title}{{MATLAB Functions for Mie Scattering and Absorption,
  Version 1}}.
\newblock \bibinfo{type}{Technical Report}.
\bibitem[{Moorhead et~al.(2017a)Moorhead, Blaauw, Moser, Campbell-Brown, Brown
  and Cooke}]{stx2175}
\bibinfo{author}{Moorhead, A.V.}, \bibinfo{author}{Blaauw, R.C.},
  \bibinfo{author}{Moser, D.E.}, \bibinfo{author}{Campbell-Brown, M.D.},
  \bibinfo{author}{Brown, P.G.}, \bibinfo{author}{Cooke, W.J.},
  \bibinfo{year}{2017}a.
\newblock \bibinfo{title}{A two-population sporadic meteoroid bulk density
  distribution and its implications for environment models}.
\newblock \bibinfo{journal}{Monthly Notices of the Royal Astronomical Society}
  \bibinfo{volume}{472}, \bibinfo{pages}{3833--3841}.
\newblock \URLprefix \url{https://doi.org/10.1093/mnras/stx2175},
  \DOIprefix\doi{10.1093/mnras/stx2175}.
\bibitem[{Moorhead et~al.(2017b)Moorhead, Cooke and
  Campbell-Brown}]{moorheadECSD}
\bibinfo{author}{Moorhead, A.V.}, \bibinfo{author}{Cooke, W.J.},
  \bibinfo{author}{Campbell-Brown, M.D.}, \bibinfo{year}{2017}b.
\newblock \bibinfo{title}{Meteor shower forecasting for spacecraft operations}.
\newblock \bibinfo{journal}{In: Proceedings of the 7th European Conference on
  Space Debris} .
\bibitem[{{Moorhead} et~al.(2019){Moorhead}, {Egal}, {Brown}, {Moser} and
  {Cooke}}]{2019JSpRo..56.1531M}
\bibinfo{author}{{Moorhead}, A.V.}, \bibinfo{author}{{Egal}, A.},
  \bibinfo{author}{{Brown}, P.G.}, \bibinfo{author}{{Moser}, D.E.},
  \bibinfo{author}{{Cooke}, W.J.}, \bibinfo{year}{2019}.
\newblock \bibinfo{title}{{Meteor shower forecasting in near-Earth space}}.
\newblock \bibinfo{journal}{Journal of Spacecraft and Rockets}
  \bibinfo{volume}{56}, \bibinfo{pages}{1531--1545}.
\newblock \DOIprefix\doi{10.2514/1.A34416}.
\bibitem[{{Narziev}(2019)}]{2019P&SS..173...42N}
\bibinfo{author}{{Narziev}, M.}, \bibinfo{year}{2019}.
\newblock \bibinfo{title}{{Physical properties of the meteoroids using
  simultaneous radar and optical observations}}.
\newblock \bibinfo{journal}{Planetary {\&} Space Science}
  \bibinfo{volume}{173}, \bibinfo{pages}{42--48}.
\newblock \DOIprefix\doi{10.1016/j.pss.2018.11.011}.
\bibitem[{Navarro and Werts(2013)}]{Navarro2013}
\bibinfo{author}{Navarro, J.R.G.}, \bibinfo{author}{Werts, M.H.V.},
  \bibinfo{year}{2013}.
\newblock \bibinfo{title}{Resonant light scattering spectroscopy of gold,
  silver and gold{\textendash}silver alloy nanoparticles and optical detection
  in microfluidic channels}.
\newblock \bibinfo{journal}{The Analyst} \bibinfo{volume}{138},
  \bibinfo{pages}{583--592}.
\newblock \URLprefix \url{https://doi.org/10.1039%2Fc2an36135c},
  \DOIprefix\doi{10.1039/c2an36135c}.
\bibitem[{{Nesvorn{\'y}} et~al.(2011){Nesvorn{\'y}}, {Janches},
  {Vokrouhlick{\'y}}, {Pokorn{\'y}}, {Bottke} and
  {Jenniskens}}]{2011ApJ...743..129N}
\bibinfo{author}{{Nesvorn{\'y}}, D.}, \bibinfo{author}{{Janches}, D.},
  \bibinfo{author}{{Vokrouhlick{\'y}}, D.}, \bibinfo{author}{{Pokorn{\'y}},
  P.}, \bibinfo{author}{{Bottke}, W.F.}, \bibinfo{author}{{Jenniskens}, P.},
  \bibinfo{year}{2011}.
\newblock \bibinfo{title}{{Dynamical Model for the Zodiacal Cloud and Sporadic
  Meteors}}.
\newblock \bibinfo{journal}{Astrophys.~J.} \bibinfo{volume}{743},
  \bibinfo{pages}{129}.
\newblock \DOIprefix\doi{10.1088/0004-637X/743/2/129},
  \href{http://arxiv.org/abs/1109.2983}{{\tt arXiv:1109.2983}}.
\bibitem[{{Pokorn{\'y}} et~al.(2014){Pokorn{\'y}}, {Vokrouhlick{\'y}},
  {Nesvorn{\'y}}, {Campbell-Brown} and {Brown}}]{2014ApJ...789...25P}
\bibinfo{author}{{Pokorn{\'y}}, P.}, \bibinfo{author}{{Vokrouhlick{\'y}}, D.},
  \bibinfo{author}{{Nesvorn{\'y}}, D.}, \bibinfo{author}{{Campbell-Brown}, M.},
  \bibinfo{author}{{Brown}, P.}, \bibinfo{year}{2014}.
\newblock \bibinfo{title}{{Dynamical Model for the Toroidal Sporadic Meteors}}.
\newblock \bibinfo{journal}{Astrophys.~J.} \bibinfo{volume}{789},
  \bibinfo{pages}{25}.
\newblock \DOIprefix\doi{10.1088/0004-637X/789/1/25}.
\bibitem[{{Protopapa} et~al.(2012){Protopapa}, {Sunshine}, {Feaga}, {Kelley},
  {Besse}, {Groussin}, {Merlin}, {Farnham}, {Li} and
  {A'Hearn}}]{2012DPS....4450601P}
\bibinfo{author}{{Protopapa}, S.}, \bibinfo{author}{{Sunshine}, J.M.},
  \bibinfo{author}{{Feaga}, L.}, \bibinfo{author}{{Kelley}, M.S.},
  \bibinfo{author}{{Besse}, S.}, \bibinfo{author}{{Groussin}, O.},
  \bibinfo{author}{{Merlin}, F.}, \bibinfo{author}{{Farnham}, T.L.},
  \bibinfo{author}{{Li}, J.}, \bibinfo{author}{{A'Hearn}, M.F.},
  \bibinfo{year}{2012}.
\newblock \bibinfo{title}{{Ice and Refractories in the Inner Coma of
  103P/Hartley 2}}, in: \bibinfo{booktitle}{AAS/Division for Planetary Sciences
  Meeting Abstracts \#44}, p. \bibinfo{pages}{506.01}.
\bibitem[{{Querry}(1985)}]{1985umo..rept.....Q}
\bibinfo{author}{{Querry}, M.R.}, \bibinfo{year}{1985}.
\newblock \bibinfo{title}{{Optical constants}}.
\newblock \bibinfo{type}{Technical Report}.
\bibitem[{{Rinaldi} et~al.(2019){Rinaldi}, {Formisano}, {Kappel}, {Capaccioni},
  {Bockel{\'e}e-Morvan}, {Cheng}, {Vincent}, {Deshapriya}, {Arnold}, {Capria},
  {Ciarniello}, {D'Aversa}, {De Sanctis}, {Doose}, {Erard}, {Federico},
  {Filacchione}, {Fink}, {Leyrat}, {Longobardo}, {Magni}, {Migliorini},
  {Mottola}, {Naletto}, {Raponi}, {Taylor}, {Tosi}, {Tozzi} and
  {Salatti}}]{2019A&A...630A..21R}
\bibinfo{author}{{Rinaldi}, G.}, \bibinfo{author}{{Formisano}, M.},
  \bibinfo{author}{{Kappel}, D.}, \bibinfo{author}{{Capaccioni}, F.},
  \bibinfo{author}{{Bockel{\'e}e-Morvan}, D.}, \bibinfo{author}{{Cheng}, Y.C.},
  \bibinfo{author}{{Vincent}, J.B.}, \bibinfo{author}{{Deshapriya}, P.},
  \bibinfo{author}{{Arnold}, G.}, \bibinfo{author}{{Capria}, M.T.},
  \bibinfo{author}{{Ciarniello}, M.}, \bibinfo{author}{{D'Aversa}, E.},
  \bibinfo{author}{{De Sanctis}, M.C.}, \bibinfo{author}{{Doose}, L.},
  \bibinfo{author}{{Erard}, S.}, \bibinfo{author}{{Federico}, C.},
  \bibinfo{author}{{Filacchione}, G.}, \bibinfo{author}{{Fink}, U.},
  \bibinfo{author}{{Leyrat}, C.}, \bibinfo{author}{{Longobardo}, A.},
  \bibinfo{author}{{Magni}, G.}, \bibinfo{author}{{Migliorini}, A.},
  \bibinfo{author}{{Mottola}, S.}, \bibinfo{author}{{Naletto}, G.},
  \bibinfo{author}{{Raponi}, A.}, \bibinfo{author}{{Taylor}, F.},
  \bibinfo{author}{{Tosi}, F.}, \bibinfo{author}{{Tozzi}, G.P.},
  \bibinfo{author}{{Salatti}, M.}, \bibinfo{year}{2019}.
\newblock \bibinfo{title}{{Analysis of night-side dust activity on comet 67P
  observed by VIRTIS-M: a new method to constrain the thermal inertia on the
  surface}}.
\newblock \bibinfo{journal}{Astronomy {\&} Astrophysics} \bibinfo{volume}{630},
  \bibinfo{pages}{A21}.
\newblock \DOIprefix\doi{10.1051/0004-6361/201834907}.
\bibitem[{{Sagan} and {Khare}(1979)}]{1979Natur.277..102S}
\bibinfo{author}{{Sagan}, C.}, \bibinfo{author}{{Khare}, B.N.},
  \bibinfo{year}{1979}.
\newblock \bibinfo{title}{{Tholins: organic chemistry of interstellar grains
  and gas}}.
\newblock \bibinfo{journal}{Nature} \bibinfo{volume}{277},
  \bibinfo{pages}{102--107}.
\newblock \DOIprefix\doi{10.1038/277102a0}.
\bibitem[{Suryanarayan(1972)}]{Suryanarayan1972}
\bibinfo{author}{Suryanarayan, E.}, \bibinfo{year}{1972}.
\newblock \bibinfo{title}{On steady parallel flows}.
\newblock \bibinfo{journal}{Journal of Mathematical Analysis and Applications}
  \bibinfo{volume}{40}, \bibinfo{pages}{791--802}.
\newblock \URLprefix \url{https://doi.org/10.1016/0022-247x(72)90020-0},
  \DOIprefix\doi{10.1016/0022-247x(72)90020-0}.
\bibitem[{{Tricarico} et~al.(2014){Tricarico}, {Samarasinha}, {Sykes}, {Li},
  {Farnham}, {Kelley}, {Farnocchia}, {Stevenson}, {Bauer} and
  {Lock}}]{2014ApJ...787L..35T}
\bibinfo{author}{{Tricarico}, P.}, \bibinfo{author}{{Samarasinha}, N.H.},
  \bibinfo{author}{{Sykes}, M.V.}, \bibinfo{author}{{Li}, J.Y.},
  \bibinfo{author}{{Farnham}, T.L.}, \bibinfo{author}{{Kelley}, M.S.P.},
  \bibinfo{author}{{Farnocchia}, D.}, \bibinfo{author}{{Stevenson}, R.},
  \bibinfo{author}{{Bauer}, J.M.}, \bibinfo{author}{{Lock}, R.E.},
  \bibinfo{year}{2014}.
\newblock \bibinfo{title}{{Delivery of Dust Grains from Comet C/2013 A1 (Siding
  Spring) to Mars}}.
\newblock \bibinfo{journal}{Astrophys.~J.l} \bibinfo{volume}{787},
  \bibinfo{pages}{L35}.
\newblock \DOIprefix\doi{10.1088/2041-8205/787/2/L35},
  \href{http://arxiv.org/abs/1404.7168}{{\tt arXiv:1404.7168}}.
\bibitem[{Wexler(1977)}]{Wexler1977}
\bibinfo{author}{Wexler, A.}, \bibinfo{year}{1977}.
\newblock \bibinfo{title}{Vapor pressure formulation for ice}.
\newblock \bibinfo{journal}{Journal of Research of the National Bureau of
  Standards Section A: Physics and Chemistry} \bibinfo{volume}{81A},
  \bibinfo{pages}{5}.
\newblock \URLprefix \url{https://doi.org/10.6028/jres.081a.003},
  \DOIprefix\doi{10.6028/jres.081a.003}.
\bibitem[{{Whipple}(1951)}]{1951ApJ...113..464W}
\bibinfo{author}{{Whipple}, F.L.}, \bibinfo{year}{1951}.
\newblock \bibinfo{title}{{A Comet Model. II. Physical Relations for Comets and
  Meteors.}}
\newblock \bibinfo{journal}{Astrophys.~J.} \bibinfo{volume}{113},
  \bibinfo{pages}{464}.
\newblock \DOIprefix\doi{10.1086/145416}.
\bibitem[{{Whipple}(1967)}]{1967SAOSR.239....1W}
\bibinfo{author}{{Whipple}, F.L.}, \bibinfo{year}{1967}.
\newblock \bibinfo{title}{{On Maintaining the Meteoritic Complex}}.
\newblock \bibinfo{journal}{SAO Special Report} \bibinfo{volume}{239},
  \bibinfo{pages}{1}.
\bibitem[{{Wiegert} et~al.(2009){Wiegert}, {Vaubaillon} and
  {Campbell-Brown}}]{2009Icar..201..295W}
\bibinfo{author}{{Wiegert}, P.}, \bibinfo{author}{{Vaubaillon}, J.},
  \bibinfo{author}{{Campbell-Brown}, M.}, \bibinfo{year}{2009}.
\newblock \bibinfo{title}{{A dynamical model of the sporadic meteoroid
  complex}}.
\newblock \bibinfo{journal}{Icarus} \bibinfo{volume}{201},
  \bibinfo{pages}{295--310}.
\newblock \DOIprefix\doi{10.1016/j.icarus.2008.12.030}.
\bibitem[{{Zook} and {Berg}(1975)}]{1975P&SS...23..183Z}
\bibinfo{author}{{Zook}, H.A.}, \bibinfo{author}{{Berg}, O.E.},
  \bibinfo{year}{1975}.
\newblock \bibinfo{title}{{A source for hyperbolic cosmic dust particles}}.
\newblock \bibinfo{journal}{Planet.~Space~Sci.} \bibinfo{volume}{23},
  \bibinfo{pages}{183--203}.
\newblock \DOIprefix\doi{10.1016/0032-0633(75)90078-1}.

\end{thebibliography}


\appendix

\section{Derivation of ejection velocity}
\label{sec:uderiv}

This appendix presents our derivation of the speed of particles carried away from a comet's nucleus by sublimation. We will assume that the sublimating gas is water. Throughout this section, we will use $v$ to denote gas speed and $u$ to denote particle speed.

\subsection{Surface temperature}

\begin{figure} \centering
\includegraphics[]{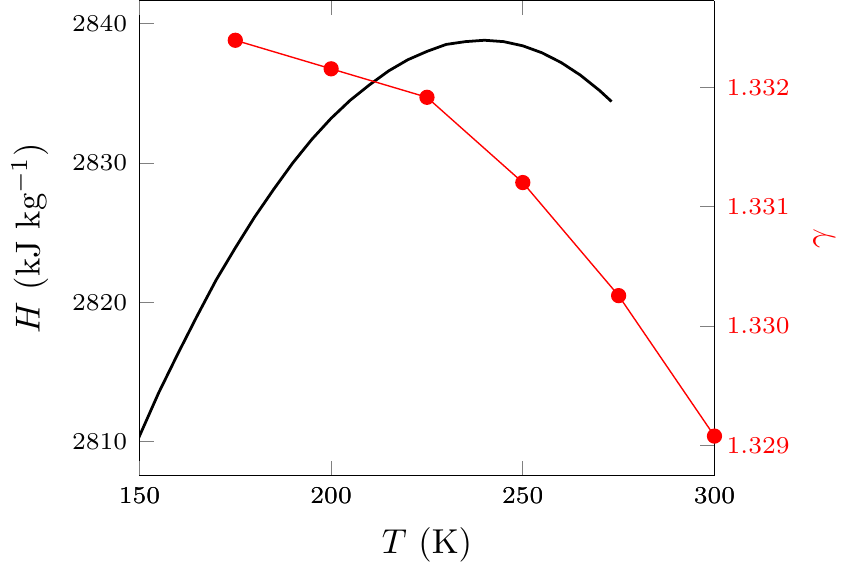}
\caption[Heat of sublimation and adiabatic index of water ice]{Heat of sublimation of water ice ($H$) and adiabatic index of water vapor ($\gamma$) as a function of temperatures, $T$. Values of $H$ are taken from Table 4 of Feistel \& Wagner (2007).  Values of $\gamma$ are computed as $\gamma = 1 - M C_p/ R$, where $M$ is the molar mass of water, $R$ is the gas constant, and $C_p$, the heat capacity at constant pressure, is taken from an online database.\footnotemark
}
\label{fig:Hg}
\end{figure}

The temperature at the surface of the comet, $T$, is governed by energy balance: the energy lost through re-radiation and sublimation must equal the solar radiation intercepted by the nucleus. Locally, the temperature will vary in a manner that depends on the solar zenith angle, geometry, rotation rate, and thermal properties of the nucleus \citep{2007Icar..187...16G}. If the heating is restricted to the surface of the nucleus, the temperature must satisfy the following energy balance equation:
\begin{align}
\frac{F}{r_h^2} \cos{\theta} &= \sigma T^4 + Z(T) \, H(T)
\label{eq:ts}
\end{align}
where $F = 1361$ W m$^{-2}$ is the solar irradiance at 1~au, $\theta$ is the local solar zenith angle, $r_h$ is heliocentric distance in au, $\sigma$ is the Stefan-Boltzmann constant, $T$ is the temperature at the surface of the comet, $Z$ is the sublimation rate, and $H(T)$ is the heat of sublimation of water ice and is shown in Fig.~\ref{fig:Hg}. 

 \footnotetext{http://www.engineeringtoolbox.com/water-vapor-d\_979.html}

The rate of sublimation is usually determined in the following manner \citep[for example, see][]{2004inco.book.....B}. One first envisions a saturated gas in equilibrium that is condensing onto a surface and vaporizing from it at the same rate, $Z$. The gas has pressure $p_\mathrm{sat}$ and molar density $n_\mathrm{sat}$ and obeys the ideal gas law:
\begin{align}
p_\mathrm{sat} &= n_\mathrm{sat} R T 
\label{eq:ideal}
\end{align}
where $R$ is the gas constant. 
The mean speed of the saturated gas is assumed to be that of a Maxwellian distribution:
\begin{align}
\bar{v}_\mathrm{sat} &= \sqrt{\frac{8 R T}{\pi M}}
\end{align}
where $M$ is the molar mass of the gas.
The rate at which particles condense is equal to the rate at which gas molecules collide with the surface:
\begin{align}
Z &= M \cdot \tfrac{1}{4} n_\mathrm{sat} \bar{v}_\mathrm{sat} = p_\mathrm{sat} \sqrt{\frac{M}{2 \pi R T}} \label{eq:zsub}
\end{align}
Note that the above formula for $Z$ contains a factor of $M$ and thus provides the mass sublimation rate in units of kg m$^{-2}$ s$^{-1}$, rather than the molar sublimation rate presented by \cite{2004inco.book.....B}. Use of a mass sublimation rate requires that the heat of sublimation, $H$, carry units of J kg$^{-1}$.

\begin{figure}
\includegraphics{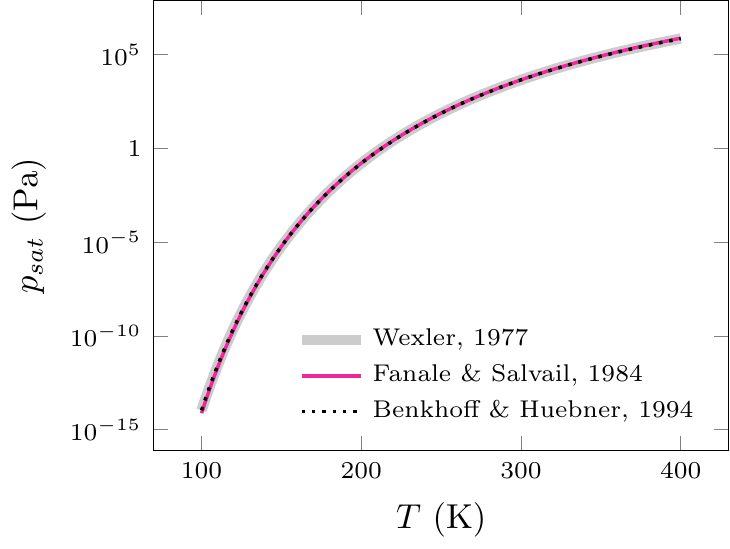}
\caption[Saturation pressure]{The pressure of saturated water vapor in equilibrium as a function of temperature.  Three different formulae are shown: Eq.~54 of \protect\cite{1995MNRAS.275..773J}, Eq.~7 of \protect\cite{1984Icar...60..476F}, and Eq.~10 of \protect\cite{1995Icar..114..348B}. The three equations agree to within 5\% between 150 and 375 K.}
\label{fig:psat}
\end{figure}

The equilibrium pressure of saturated water vapor is a function of temperature; several works provide expressions for the saturation pressure as a function of temperature \citep{Wexler1977,1984Icar...60..476F,1995Icar..114..348B}.
The pressure quoted by these three papers is extremely similar between 100 and 400~K (see Fig.~\ref{fig:psat}) and so we opt to use the simplest equation of the three \citep{1984Icar...60..476F}:
\begin{align}
p_\mathrm{sat} &= (3.56 \times 10^{12} \mbox{ Pa}) \, \cdot e^{-6141.667\mathrm{~K}/T}
\label{eq:psat}
\end{align}

\begin{figure}
\includegraphics{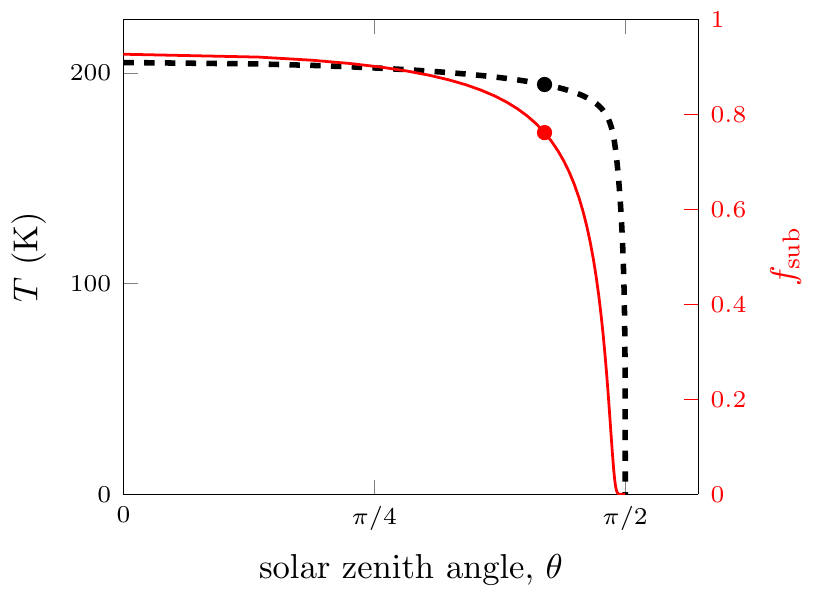}
\caption{Comet nucleus surface temperature (black, dashed) and the fraction of the incident energy that goes into sublimation (red) as a function of solar zenith angle, $\theta$. The dots mark the angle where $\cos \theta = \tfrac{1}{4}$ and the temperature is equivalent to that of an evenly heated nucleus.}
\label{fig:temp}
\end{figure}

The combination of Eqs.~\ref{eq:ts}, \ref{eq:zsub}, and \ref{eq:psat} allows us to solve for the surface temperature of the comet as a function of heliocentric distance and solar zenith angle:
\begin{align}
\frac{F}{r_h^2} \cos{\theta} 
&= \sigma T^4 + H(T) \cdot p_\mathrm{sat}(T) \cdot \sqrt{\frac{M}{2 \pi R T}}
\label{eq:tcombo}
\end{align}
Figure \ref{fig:temp} presents temperature as a function of solar zenith angle, assuming a heliocentric distance of 1~au. At the subsolar point, the surface temperature is 204.9~K, which is very close to the value of 204.5 presented by \cite{1995MNRAS.275..773J}. Figure~\ref{fig:temp} also presents the fraction of the energy that goes into sublimation rather than re-radiation ($f_\mathrm{sub}$); close to the terminator, sublimation plummets and radiation dominates.

Equation~\ref{eq:ts} applies if heating of the cometary surface is local. If we instead assume that the nucleus is evenly heated and sublimation takes place evenly across the entire surface of the nucleus, the energy balance equation becomes:
\begin{align}
\frac{F}{4 r_h^2} &= \sigma T^4 + Z(T) \, H(T) \label{eq:tn}
\end{align}
where the factor of 4 arises from the ratio of a sphere's cross-sectional area to its surface area. Note that Eq.~\ref{eq:tn} and \ref{eq:ts} are equal when $\cos \theta = \tfrac{1}{4}$, which occurs when $\theta \simeq 1.32$ radians. The values of $T$ and $f_\mathrm{sub}$ (194.5$^\circ$~K and 0.76, respectively) that correspond to this angle are marked in Fig.~\ref{fig:temp}.

\subsection{Gas dynamics}

Now that we have determined the temperature of the surface of the nucleus, we will turn our attention to the motion of the gas as it streams away from the comet.

\subsubsection{Gas dynamics at the comet's surface}

We will assume that the density and pressure of the gas are initially half their saturation values \citep{2004Icar..168..163D}. This corresponds to a scenario in which we envision a saturated gas that is both condensing onto a surface and sublimating from it at a rate $Z$, and then discard the condensing component, obtaining a factor of 2 reduction in number density ($n_0 = \tfrac{1}{2} n_\mathrm{sat}$). Only outward-moving particles remain and therefore the gas is not in equilibrium \citep{2004Icar..168..163D}. Note that this simplifying assumption of no condensation is not strictly true; \cite{2004Icar..168..163D} estimate the backflux as varying between 4\% and 22\%. 

The average speed of the water molecules is unchanged, but only outward-bound molecules remain. Thus, the average {\sl outward} speed of the gas near the surface of the comet is given by:
\begin{align}
v_0 &= \int_0^{\pi/2} \bar{v}_\mathrm{sat} \cos{\theta} \sin{\theta} ~ d\theta 
     = \tfrac{1}{2} \bar{v}_\mathrm{sat} \label{eq:v0}
\end{align}
This speed appears in, among others, \cite{2002MNRAS.337.1081M}. The sublimation rate is $Z = M n_0 v_0$; mass is conserved regardless of whether the gas is in thermodynamic equilibrium.

\subsubsection{Re-establishment of thermal equilibrium}

\cite{2000Icar..148..594H} derived the temperature and bulk velocity of a sublimating gas after it has re-established thermodynamic equilibrium. The final temperature of the gas depends on the adiabatic gas constant and the number of excited rotational and vibrational degrees of freedom, but the final bulk velocity can be expressed fairly simply in terms of the {\sl initial} average speed:
\begin{align}
v_1 &= \frac{\pi \bar{v}_{sat}}{4} = \sqrt{\frac{\pi R T}{2 M}} \label{eq:v1a}
\end{align}
where $T$ is the initial temperature of the gas and of the comet's surface. Note that Eq.~\ref{eq:v1a} is equivalent to Eq.~4 of \cite{1995MNRAS.275..773J}. This velocity is larger than $v_0$ because some of the initial thermal energy of the gas has been converted to bulk velocity. Once the bulk velocity has been determined, we can use $Z = \rho_1 v_1$ to obtain the mass density of the gas, or 
\begin{align}
\rho_1 &= \frac{p_\mathrm{sat}(T)}{v_1} \sqrt{\frac{M}{2 \pi R T}} = p_\mathrm{sat}(T) \frac{M}{\pi R T}
\label{eq:rho1}
\end{align}

We assume that both the re-establishment of equilibrium and the subsequent outward flow of the gas are adiabatic processes. Thus, since some of the initial thermal energy of the gas is converted to bulk motion, the temperature of the gas must be lower than the temperature of the comet's surface. The new temperature is given by \cite{2000Icar..148..594H} as:
\begin{align}
T_1 &= \frac{8 + 2 f_{rv} - \pi}{2(f_{rv}+3)}T
\end{align}
where $f_{rv}$ is the number of excited rotational and vibration degrees of freedom in the gas. The total number of degrees of freedom is $f = f_{rv} + 3$. If we substitute $f = 2/(\gamma-1)$, where $\gamma$ is the adiabatic index (see Fig.~\ref{fig:Hg}), we obtain:
\begin{align}
T_1 &= \tfrac{1}{4} (2 + \pi + 2 \gamma - \pi \gamma) \, T \label{eq:t1}
\end{align}

The above equations from \cite{2000Icar..148..594H} require a number of collisions to occur in order for the gas to reestablish equilibrium. The gas does not achieve equilibrium instantly, but rather several mean free paths distant from the surface. The mean free path is approximately
\begin{align}
l &\sim \frac{1}{\sqrt{2} \pi s_K^2 n N_A} \label{eq:lpath}
\end{align}
where $s_K$ is the kinetic radius of the gas molecule and $N_A$ is Avogadro's  number. Using 265 pm for the kinetic radius of a water molecule \citep[see Table 2.2 of][]{Ismail2015} and calculating $n_1$ at $r = 1$ au and $\theta = 0^\circ$, we obtain a mean free path of 0.083~m. Thus, one meter could be considered ``many collisional mean free paths away from the surface" \citep{2000Icar..148..594H}. One meter is also generally negligible in comparison with cometary nuclei measuring $\sim 1$ km in diameter, and so we will henceforth use Eq.~\ref{eq:v1a} to compute the gas velocity ``at the surface of the nucleus.''

\subsubsection{Outward gas flow}

As the gas expands and flows away from the cometary surface, it follows the relation \citep{Suryanarayan1972,1995MNRAS.275..773J}:
\begin{align}
\frac{d p}{d s_f} + \rho v \frac{d v}{d s_f} &= 0 \label{eq:flow}
\end{align}
where $s_f$ is distance along the flow line. Since the gas is assumed not to exchange energy with the environment, it should also obey the adiabatic
gas law:
\begin{align}
p \rho^{-\gamma} &= p_1 \rho_1^{-\gamma} = C
\end{align}
where $\gamma$ is the adiabatic index, C is a constant, and the subscript 1 indicates that the values are taken near the surface of the comet but after thermal equilibrium has been re-established. If we substitute $p = C \rho^{\gamma}$ into Eq.~\ref{eq:flow} and divide by $\rho$, we obtain
\begin{align}
C \, \gamma \, \rho^{\gamma-2} \frac{d \rho}{d s_f} + v \frac{d v}{d s_f} &= 0
\end{align}
The above equation can be integrated and multiplied by 2 to obtain
\begin{align}
\frac{2 C \gamma}{\gamma - 1} \rho^{\gamma - 1} + v^2 &= D \label{eq:d}
\end{align}
where $D$ is another constant. Thus, we must determine the values of the constants $C$ and $D$.

The equilibrium pressure near the surface can be calculated from the ideal gas law:
\begin{align}
p_1 &= \frac{\rho_1}{M} R T_1 \label{eq:ideal2}
\end{align}
where the temperature $T_1$ is given by Eq.~\ref{eq:t1}. If we further substitute $T = 2 M v_1^2/\pi R$, the pressure becomes
\begin{align}
p_1 &= \frac{2 + \pi + 2 \gamma - \pi \gamma}{2 \pi} \, \rho_1 v_1^2 \label{eq:p1v1}
\end{align}
and therefore 
\begin{align}
C &= p_1 \rho_1^{-\gamma} = 
\frac{2 + \pi + 2 \gamma - \pi \gamma}{2 \pi} \, \rho_1^{1-\gamma} v_1^2
\end{align}
We can now insert this relation for $C$ into Eq.~\ref{eq:d} and evaluate at location 1:
\begin{align}
D &= \frac{\gamma}{\gamma - 1} \frac{2 + \pi + 2 \gamma - \pi \gamma}{\pi} 
\left(\frac{\rho}{\rho_1}\right)^{\gamma-1}\, v_1^2 + v^2 
\end{align}
Now that our constants have been determined, we can relate velocity and density along the flow line using an expression that is similar to Eq.~7 of \cite{1995MNRAS.275..773J}:
\begin{align}
v_n^2 &= 1 + g \, \left( 1 - \rho_n^{\gamma - 1} \right)
\label{eq:vrho} \\
g &= \frac{\gamma}{\gamma - 1} \frac{2 + \pi + 2 \gamma - \pi \gamma}{\pi} \label{eq:g}
\end{align}
Note that we have expressed the gas velocity and density relative to their values ``at the surface'': $v_n = v/v_1$ and $\rho_n = \rho/\rho_1$. We have also introduced the dimensionless quantity $g$. \cite{1995MNRAS.275..773J} obtains $g = 4 \gamma / \pi (\gamma - 1)$; this corresponds to the assumption that $T_1 = T$, which is not correct according to \cite{2000Icar..148..594H}. Nevertheless, equation~\ref{eq:g} yields a value that is similar to the simpler expression from \cite{1995MNRAS.275..773J}; for $\gamma = 1.332$, we obtain $g = 4.62$ rather than 5.11. \cite{1995MNRAS.275..773J} actually quotes a value of 5.38, but the author confirmed that this was an error (J.\ Jones, personal comm.).

As noted by \cite{1995MNRAS.275..773J}, the mass flux along the flow line is the product of $\rho$ and $v$:
\begin{align}
J &= \rho v \label{eq:Js}
\end{align}

\begin{figure}
\includegraphics{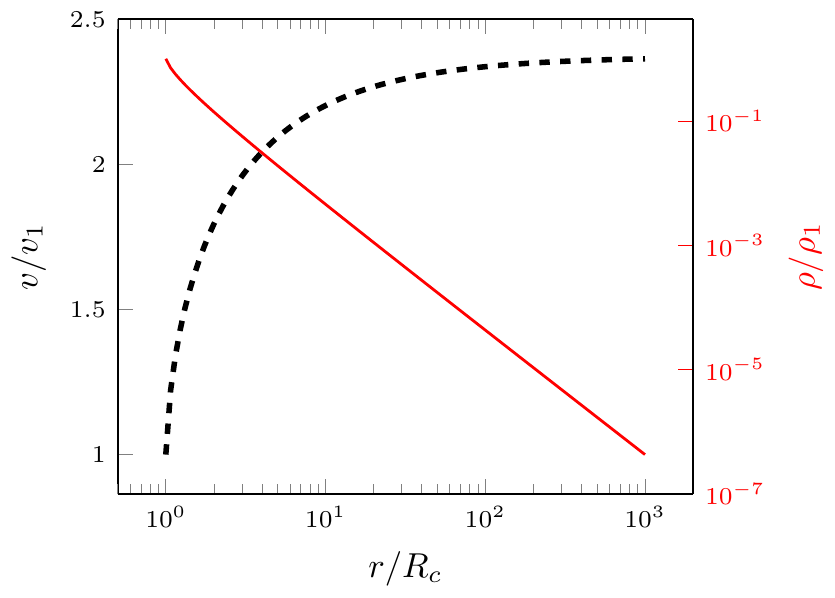}
\caption[Bulk velocity and mass density as a function of cometocentric distance]{Bulk velocity, $v$, and mass density, $\rho$, of gases sublimating from a comet relative to their values near the surface of the comet ($v_1$, $\rho_1$) and as a function of cometocentric distance ($r$). These solutions correspond to the spherically symmetric case.}
\label{fig:gas}
\end{figure}

Let us momentarily neglect the dependence of Eq.~\ref{eq:tcombo} on solar zenith angle, and assume that the temperature, pressure, outward bulk velocity, and gas density is the same across the entire surface of the nucleus. In this case, the gas will flow directly outward at all points and the mass flux will therefore obey:
\begin{align}
\rho_n v_n &= (r/R_c)^{-2} = r_n^{-2}
\label{eq:jsphere}
\end{align}
where $r$ is cometocentric distance and $R_c$ is the radius of the nucleus. Eq.~\ref{eq:jsphere}, when combined with Eq.~\ref{eq:vrho}, can be solved to obtain the velocity and density of the gas as a function of distance from the comet (see Fig.~\ref{fig:gas}).

When the outgassing is not spherically symmetric, one must solve for both the radial and tangential components of the mass flux. \cite{1995MNRAS.275..773J} uses Legendre polynomials to describe an axisymmetric flow, following the general approach of \cite{JacksonEM}. 
However, the Legendre polynomial approach is typically solved by imposing a boundary condition on the potential, $\Phi$ \citep{JacksonEM}. \cite{1995MNRAS.275..773J} instead imposes a boundary condition on the radial component of the gradiant of the potential, $J_r$. We find that his solution correctly reproduces $J_r$, but introduces a strong tangential flow at the surface of the comet, $J_\theta$. The vector sum of these two components exceeds $v_1$ and is, we believe, incompatible with our assumptions. We plan to revisit the axisymmetric case in the future in an attempt to resolve this issue. In this paper, we will confine ourselves to the spherically symmetric case.

\subsection{Particle dynamics}

Once the gas flow has been determined, we can solve for the motion of solid particles due to gas drag:
\begin{align}
\frac{d^2 r}{d t^2} &= \frac{d u}{d t} = \frac{A \Gamma}{2} m^{-1/3} \rho_m^{-2/3} \rho \, [v - u]^2 \label{eq:dudt}
\end{align}
where $u$ is the radial speed of the particle and $r$ its cometocentric distance, $m$ is the particle's mass, and $\rho_m$ is the particle's bulk density.  
The parameter $A$ is the shape factor, which describes the ratio of a particle's cross-sectional area to its volume raised to the 2/3 power. We assume that the ejected particles are spheres and that $A = (9 \pi / 16)^{1/3} \simeq 1.209$, but irregular particles may have higher shape factors and thus be accelerated faster. The parameter $\Gamma$ is the drag coefficient; this work follows \cite{1951ApJ...113..464W} and \cite{1995MNRAS.275..773J} in using $\Gamma = 2(1 +  \tfrac{4}{9}) \simeq 2.89$. This value corresponds to free molecular flow, in which each colliding gas molecule is completely stopped by the meteoroid, then re-emitted at thermal velocity in a random direction. Finally, we assume that $r = R_c$ and $u = 0$ at the surface of the nucleus.

To simplify this process, we define dimensionless variables $r_n = r/R_c$, $u_n = u/v_1$, and $t_n = t \cdot v_1/R_c$, and a dimensionless constant $\xi$, where
\begin{align}
\xi^2 &= \frac{A \Gamma}{2} m^{-1/3} \rho_m^{-2/3} \rho_1 \, R_c \label{eq:xi}
\end{align}
Thus, our system of equations becomes:
\begin{align}
\frac{d u_n}{d t_n} &= \xi^2 \, \rho_n \, \left[{v_n - u_n }\right]^2 \label{eq:dundt} \\
\frac{d r_n}{d t_n} &= u_n \label{eq:dxndt}
\end{align}
The parameter $\xi$ encompasses all dependence on particle mass, bulk particle density, particle shape factor, comet size, and heliocentric distance.  The dimensionless parameters $\rho_n$ and $v_n$ depend only on normalized cometocentric distance $r_n$ and the adiabatic index $\gamma$.  Although the adiabatic index $\gamma$ is technically a function of temperature and thus heliocentric distance, it varies so little over our range of interest (see Fig.~\ref{fig:Hg}) that we treat it as a constant.

\subsubsection{Large particles ($\xi \ll 1$)}
\label{sec:smallxi}

\begin{figure} \centering
\includegraphics[]{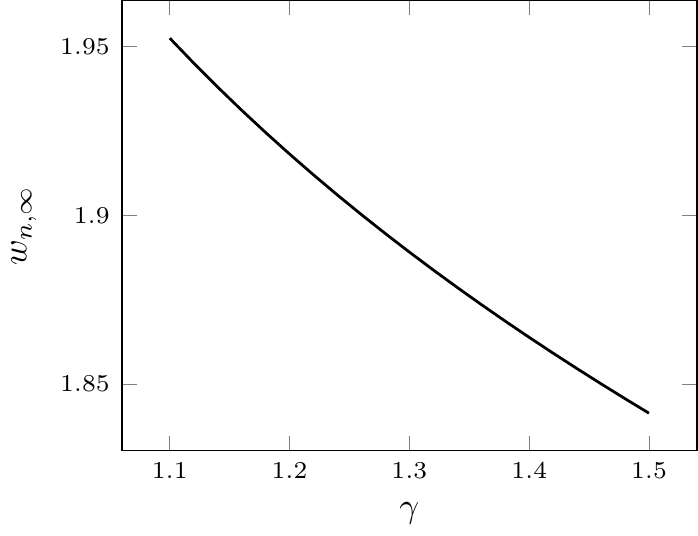}
\caption{Velocity parameter $w_n$ at arbitrarily large cometocentric distance as a function of adiabatic index $\gamma$. Note that $w_{n,\infty}$ is not particularly sensitive to modest variations in $\gamma$. }
\label{fig:gv}
\end{figure}

When $\xi \ll 1$, $u_n$ remains small compared to $v_n$ and may be dropped from Eq.~\ref{eq:dundt}.  This is the assumption that \cite{1995MNRAS.275..773J} makes in order to solve Eq.~\ref{eq:dudt} for large particles. If we further substitute $w_n = u_n/\xi$ and $s_n = t_n \xi$, we find:
\begin{align}
\frac{d w_n}{d s_n} &= \rho_n \, v_n^2 \\
\frac{d x_n}{d s_n} &= w_n
\end{align}
If we integrate these equations to obtain $w_n$ at large cometocentric distance, we obtain $w_{n,\infty} = 1.881$; this result is independent of the value of $\xi$. The value of $w_{n,\infty}$ does vary with $\gamma$, but Fig.~\ref{fig:gv} shows that these variations are modest.

We now can obtain the particle velocity at large cometocentric distance as follows:
\begin{align}
u_{\infty,a} &=  v_1 \, \xi \, w_{n,\infty} \label{eq:ufbig}
\end{align}

\subsubsection{Small particles ($\xi \gg 1$)}
\label{sec:bigxi}

When $\xi$ is large, the dust particles rapidly accelerate until $u \sim v$.  Thus, we can in this case simply set 
\begin{align}
u_{\infty,b} &= v_\infty = v_1 \sqrt{1 + g} \label{eq:ufsmall}
\end{align}
where the parameter $g$ is defined in Eq.~\ref{eq:g}. This expression does not vary with particle size or bulk density. 

\subsubsection{All particle sizes}

\begin{figure} \centering
\includegraphics[]{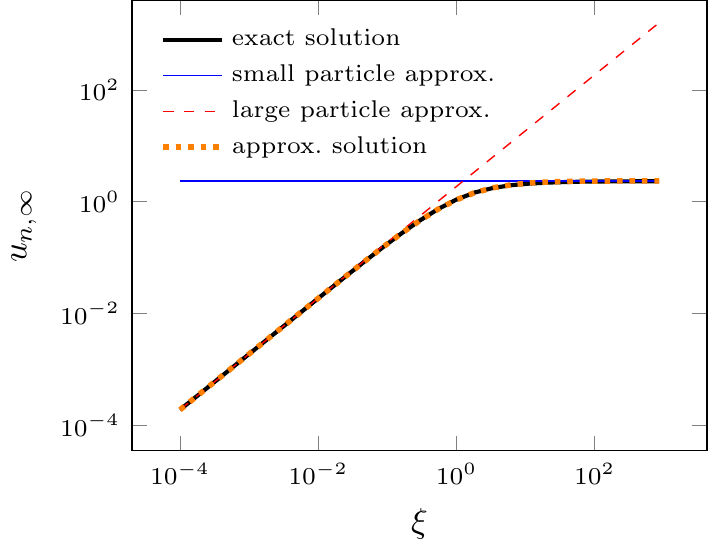}
\caption{The final velocity of dust particles carried outward from a cometary nucleus by sublimating water ice.  We present a unitless normalized velocity; the true velocity $u_\infty = v_1 u_{n,\infty}$, where $v_1$ is the outward speed of the sublimating gas near the comet's surface.  Velocity is shown as a function of the dimensionless parameter $\xi$, which depends on physical parameters of the dust particle and comet nucleus (see Eq.~\ref{eq:xi}).  We display the exact solution for $u_{n,\infty}$, the two limiting cases discussed here, and the approximate solution given by Eq.~\ref{eq:uapprox}.}
\label{fig:wvu}
\end{figure}

Figure \ref{fig:wvu} shows the normalized dust speed at large cometocentric distance for varying values of $\xi$.  Note that for small $\xi$ (large particles), $u_{n,\infty}$ approaches the large particle approximation discussed in Section \ref{sec:smallxi}, while for large $\xi$ (small particles), $u_{n,\infty}$ approaches the small particle approximation discussed in Section \ref{sec:bigxi}.

For a more rapid calculation of $u_{n,\infty}$, we provide an empirically-determined relation that combines our two limiting cases. We multiply by $v_1$ to obtain the dimensional speed $u_\infty$:
\begin{align}
u_\infty &\simeq v_1(T) \left({
(\sqrt{1+g})^{-1.054} + (w_{n,\infty} \xi)^{-1.054}}\right)^{-0.949} \nonumber \\
&\simeq v_1(T) \left({0.4025 + 0.5139 \cdot \xi^{-1.054}}\right)^{-0.949}
\label{eq:uapprox}
\end{align}
for $\gamma = 1.332$. The speed $v_1 (T)$ can be calculated using Eqs.~\ref{eq:tn} and \ref{eq:v1a}, and $\xi$ can be computed using Eqs.~\ref{eq:rho1} and \ref{eq:xi}. We will use Eq.~\ref{eq:uapprox} to compute the final outward speed of meteoroids and dust particles from this point forward.

\section{An exact solution for unbound large particles}
\label{sec:large}

In this section, we present an exact analytic solution for the mass at which particles become unbound, assuming that geometric optics applies and that the particles' ejection velocity follows the large particle limit. 

\subsection{Leading ejecta}

We first consider the case in which particles are ejected in the direction of the comet's motion.  In this case, the particle's velocity is the sum of the comet's velocity and ejection velocity.  Such particles are unbound when:
\begin{align}
v_p + u_\infty(\beta) &= v_{esc}(\beta)
\end{align}
For the geometric optics case, $\beta \propto (\rho s)^{-1} \propto m^{-1/3} \rho^{-2/3}$.  Thus, $\beta$ has the same dependence on particle mass and density as $\xi^2$.  Using this fact, we make the $\beta$ dependence explicit:
\begin{align}
    v_p + A \sqrt{\beta} &= B \sqrt{1-\beta}
\end{align}
where $A = u_\infty(\beta=1)$ and $B = v_{esc}(\beta=0)$.  We can further rearrange this as:
\begin{align}
- A \sqrt{\beta} + B \sqrt{1-\beta} &= v_p \label{eq:simp1}
\end{align}
Next, if we substitute $\sin y = -\sqrt{\beta}$, we obtain:
\begin{align}
A \sin y + B \cos y &= v_p \label{eq:sub1}
\end{align}
which has the solution
\begin{align}
y &= \sin^{-1} \left({\frac{v_p}{\sqrt{A^2 + B^2}}}\right) - \mbox{atan2}(B,A) \label{eq:y1} \\
\beta_1 &= \sin^2 y \label{eq:b1}
\end{align}

Equations \ref{eq:simp1} and \ref{eq:sub1} are equivalent only when $\sin y$ is negative and $\cos y$ is positive.  We therefore next verify that these conditions are satisfied.  We note that $z = u_\infty/v_{esc}$ is much less than unity, and that $v_p/v_{esc}(\beta=0) = \sqrt{1-q/2a}$.  Furthermore, both $v_{esc}$ and $u_\infty$ are always positive and therefore atan2$(B,A) = \tan^{-1}(B/A)$.
Using these relations, we can express Eq.~\ref{eq:y1} as
\begin{align}
y &= \sin^{-1} \left({\sqrt{\frac{1-q/2a}{1+z^2}}}\right)
- \tan^{-1}\left({\frac{1}{z}}\right) \nonumber\\
&= \sin^{-1} \left({\frac{\sqrt{1-q/2a}}{\sqrt{1+z^2}}}\right)
- \sin^{-1}\left({\frac{1}{\sqrt{1+z^2}}}\right)
\end{align}
For periodic comets, $q/2a$ lies between 0 and 1 (as does $z$).  Thus both arguments of $\sin^{-1}$ in the above equation also lie between 0 and 1, although 
\begin{align}
{\frac{1}{\sqrt{1+z^2}}} > {\frac{\sqrt{1-q/2a}}{\sqrt{1+z^2}}}
\end{align}
The function $\sin^{-1}$ is monotonically increasing over the interval (0,1), with a maximum value of $\pi/2$ and a minimum of 0.  Thus, $y$ is less than zero but greater than $-\pi/2$, and therefore satisfies $\sin y < 0$ and $\cos y > 0$.

\subsection{Trailing ejecta}

Next, we consider the case in which particles are ejected opposite to the direction of the comet's motion.  In this case, the ejection velocity must be subtracted from the comet's velocity, and the resulting orbits are unbound when:
\begin{align}
    v_p - A \sqrt{\beta} &= B \sqrt{1-\beta}
\end{align}
or
\begin{align}
A \sqrt{\beta} + B \sqrt{1-\beta} &= v_p \label{eq:simp2}
\end{align}
In this case we make the substitution $\cos x = \sqrt{\beta}$ and obtain:
\begin{align}
A \cos x + B \sin x &= v_p \label{eq:sub2}
\end{align}
which has the solution:
\begin{align}
x &= \sin^{-1} \left({\frac{v_p}{\sqrt{A^2 + B^2}}}\right) - \mbox{atan2}(A,B) \label{eq:x2} \\
\beta_2 &= \cos^2 x \label{eq:b2}
\end{align}

Equations \ref{eq:simp2} and \ref{eq:sub2} are equivalent only when $\sin x$ and $\cos x$ are both positive.  In this case, we obtain: 
\begin{align}
x &= \sin^{-1} \left({\sqrt{\frac{1-q/2a}{1+z^2}}}\right)
- \tan^{-1}(z) \nonumber \\
 &= \sin^{-1} \left({\sqrt{\frac{1-q/2a}{1+z^2}}}\right)
- \sin^{-1}\left({\frac{z}{\sqrt{1+z^2}}}\right)
\end{align}
So long as the comet's orbit is eccentric enought to satisfy $\sqrt{1 - q/2a} > z$, $x$ lies in the first quadrant and satisfies $\sin y > 0$ and $\cos y > 0$.

Equations \ref{eq:b1} and \ref{eq:b2} provide analytic solutions for the critical $\beta$ value for particles leading and trailing the comet.  These equations should be applicable for all formulations for ejection velocity that are proportional to $\sqrt{\beta}$.

\end{document}